\documentclass{statsoc}

\usepackage[a4paper, left=20mm, bottom=30mm]{geometry}
\usepackage{graphicx}
\usepackage[textwidth=10em,textsize=small]{todonotes}
\usepackage{amsmath, bbm, amssymb}
\usepackage{natbib}
\usepackage{multirow}
\usepackage{adjustbox}
\usepackage{comment}
\usepackage{hyperref}
\usepackage{etoolbox}
\usepackage[shortlabels]{enumitem}
\usepackage{tabularx}
\usepackage{booktabs}
\usepackage{algorithm}
\usepackage{algpseudocode}

\makeatletter
\patchcmd{\@makecaption}
  {\parbox}
  {\advance\@tempdima-\fontdimen2} 
  {}{}
\makeatother 

\title[]{Optimal Individualized Treatment Rule For Combination Treatments Under Budget Constraints}
\author{Qi Xu}
\address{University of California Irvine,
Irvine,
USA.}
\email{qxu6@uci.edu}
\author{Haoda Fu}
\address{Eli Lilly and Company,
Indianapolis,
USA.}
\email{fu\_haoda@lilly.com}
\author[Xu et al.]{Annie Qu}
\address{University of California Irvine,
Irvine,
USA.
}
\email{aqu2@uci.edu}

\newtheorem{prop}{Proposition}

\newtheorem{lemma}{Lemma}
\newtheorem{theorem}{Theorem}
\newtheorem{corollary}{Corollary}
\newtheorem{assumption}{Assumption}

\DeclareMathOperator*{\argmin}{argmin}
\DeclareMathOperator*{\argmax}{argmax}

\newcommand{\indep}{\perp \!\!\! \perp}

\begin{document}

\begin{abstract}

The individualized treatment rule (ITR), which recommends an optimal treatment based on individual characteristics, has drawn considerable interest from many areas such as precision medicine, personalized education, and personalized marketing. Existing ITR estimation methods mainly adopt one of two or more treatments. However, a combination of multiple treatments could be more powerful in various areas. In this paper, we propose a novel Double Encoder Model (DEM) to estimate the individualized treatment rule for combination treatments. The proposed double encoder model is a nonparametric model which not only flexibly incorporates complex treatment effects and interaction effects among treatments, but also improves estimation efficiency via the parameter-sharing feature. In addition, we tailor the estimated ITR to budget constraints through a multi-choice knapsack formulation, which enhances our proposed method under restricted-resource scenarios. In theory, we provide the value reduction bound with or without budget constraints, and an improved convergence rate with respect to the number of treatments under the DEM. Our simulation studies show that the proposed method outperforms the existing ITR estimation in various settings. We also demonstrate the superior performance of the proposed method in PDX data that recommends optimal combination treatments to shrink the tumor size of the colorectal cancer.

\end{abstract}

\section{Introduction}

Individualized decision-making has played a prominent role in many fields such as precision medicine, personalized education, and personalized marketing due to the rapid development of personalized data collection. For example, in precision medicine, individualized treatments based on individuals' demographic information and their overall comorbidity improve healthcare quality \citep{schmieder2015achievement}. However, most existing individualized decision-making approaches select one out of multiple treatments, whereas recent advances in medical and marketing research have suggested that applying multiple treatments simultaneously, referred to as combination treatments, could enhance overall healthcare. Specifically, combination treatments are able to reduce treatment failure or fatality rates, and overcome treatment resistance for many chronic diseases \citep[e.g.,][]{mokhtari2017combination, kalra2010combination, maruthur2016diabetes, bozic2013evolutionary, korkut2015perturbation, mottonen1999comparison, forrest2010rifampin, tamma2012combination}. Therefore, it is critical to develop a novel statistical method to recommend individualized combination treatments.

There are various existing methods for estimating the optimal individualized treatment rule. The first approach is the model-based approach, which estimates an outcome regression model given pre-treatment covariates and the treatment. The optimal ITR is derived by maximizing the outcome over possible treatments conditioned on the pre-treatment covariates. Existing works such as Q-learning \citep{qian2011performance}, A-learning \citep{lu2013variable, shi2018high} and RD-learning \citep{meng2020doubly} all belong to this approach. The other approach is known as the direct-search approach, which directly maximizes the expected outcome over a class of decision functions to obtain an optimal ITR. The seminal works of the direct-search approach include outcome weighted learning \citep{zhao2012estimating, huang2019multicategory}, residual weighted learning \citep{zhou2017residual}, and augmented outcome weighted learning \citep{zhou2017augmented}. However, the aforementioned methods in these two categories are designed for selecting one optimal treatment among two or more treatments. In order to accommodate combination treatments, \citet{liang2018estimating} proposed an outcome weighted learning approach using the Hamming loss, extending the direct-search approaches to estimate optimal individualized treatment rules for combination treatments. Except for outcome weighted learning with the Hamming loss \citep{liang2018estimating}, the other methods treat each combination treatment as independent ones, ignoring the correlation between different combinations. Consequently, this type of modeling strategy suffers from the curse of dimensionality issue due to the combinatorial nature of combination treatments, accompanied by computation cost increases and estimation efficiency sacrifices. In addition, the method in \citep{liang2018estimating} ignores the interaction effects among different treatments, which leads to inconsistent estimation of the ITR \citep{liang2018estimating}. In summary, existing methods ignore either correlation among combinations or interactions among treatments, yet both of them are essential to ensure an accurate and efficient estimation of the individualized treatment rule for combination treatments.

In this paper, we propose a double encoder model (DEM) to estimate the optimal individualized treatment rule for combination treatments. The proposed method incorporates both the interaction effects among different treatments and correlations among different combinations. Specifically, we introduce an outcome regression model where the treatment effects are represented by the inner product between the pre-treatment covariates and the treatment encoder. Specifically, the treatment encoder is decoupled as the additive treatment encoder and the interactive treatment encoder, where the interactive treatment encoder explicitly models the interaction effects of combination treatments. Meanwhile, the covariates encoder allows either parametric or nonparametric models to learn a low-dimensional representation of pre-treatment covariates. Finally, we derive the optimal individualized treatment rule for combination treatments by maximizing the outcome regression model over the combination treatments. As we developed our method, a parallel work \citep{kaddour2021causal} proposed the generalized Robinson decomposition, which estimates the conditional average treatment effects (CATE) for structured treatments such as graphs, images, and texts. Their proposed generalized Robinson decomposition also utilizes two neural networks to represent the treatment effects given covariates $\mathbf{X}$ and treatments $\mathbf{A}$. In spite of the overlap, our proposed method targets the combination treatments, especially considering the interaction effects among different treatments and correlations between different combinations.

Furthermore, the combination treatments assignment might be restricted by limited resources in a real world scenario. Existing works \citep{luedtke2016optimal, kitagawa2018should} consider the total amount constraint for binary treatments only, where the assignments are determined by the quantiles of treatment effects. In contrast, allocating combinations of treatments with a limited amount is an NP-hard problem, thus an analytical solution like quantiles does not exist. To address these problem, we formulate the constrained individualized treatment rule as a multi-choice knapsack problem \citep{kellerer2004multidimensional}, and solve this optimization problem through an efficient dynamic programming algorithm.

The main advantages and contributions of this paper are summarized as follows. First of all, the proposed method addresses the curse of dimensionality issue in combination treatment problems through a double encoder framework. In this framework, the covariates encoder captures the shared function bases of treatment effects, while the treatment encoder learns the coefficients for those function bases. This approach shifts complexity of treatment effects to the complexities of the covariates and treatment encoder, which are managed by our well-designed structural architecture. Second, the proposed method enhances the estimation efficiency and the rate of value reduction convergence through the parameter-sharing feature inherent in the neural network-based treatment encoder. Third, the nonparametric modeling strategy employed by the double encoder model accommodates the intricate treatment and interaction effects, effectively mitigating the model mis-specification problem and leading to consistent estimation of the ITR. Fourth, the proposed multi-choice knapsack framework enables the tailoring of individualized decisions within budget constraints. Apart from its application in the proposed method, this framework is generally applicable to other outcome regression models for deriving budget-constrained decisions for multi-arm or combination treatment scenarios.

In regards to the theoretical properties of the estimated ITR, we provide the value reduction bound for the ITR for combination treatments with or without budget constraints. Thereafter, we provide a non-asymptotic value reduction bound for the DEM, which guarantees that the value function of the estimated individualized treatment rule converges to the optimal value function with a high probability and the proposed method achieves a faster convergence rate compared with existing methods for the multi-arm ITR. The improvement in convergence rate is attained by the hierarchical structure of the neural network where the parameters are shared by all combinations and the input dimension is proportional to the number of treatments instead of the total number of combination treatments.

The proposed method demonstrates superior performance over existing methods in our numerical studies especially when the number of treatments is large and there exist interaction effects among different treatments. In the real data application, we apply the proposed method to recommend the optimal combination treatments to shrink tumor size of colorectal cancer, which achieves the maximal tumor size shrinkage and shows its potential in improving individualized healthcare.

The rest of this paper is organized as follows. In Section 2, we introduce the notations and background of the Q-learning framework, and the budget constraints problem. In Section 3, we propose the double encoder model (DEM) to estimate the optimal individualized treatment rule for combination treatments and impose a budget constraint on the original problem. In Section 4, we establish the theoretical properties of the proposed method. In Section 5, we illustrate the empirical performance of the proposed method in various simulation studies. In Section 6, we apply the proposed method to PDX data, which aims to recommend optimal combination treatments for colorectal cancer. We provide discussion and concluding remarks in Section 7.

\section{Notations and Background}
In this section, we introduce the problem setup and notations for the estimation of individualized treatment rules for combination treatments. Consider the data $(\mathbf{X}, \mathbf{A}, Y)$ collected from designed experiments or observational studies. The subject pre-treatment covariates are denoted by $\mathbf{X} \in \mathcal{X} \subset \mathbb{R}^p$, which might include patients' demographics and lab test results. The combinations of $K$ treatments are denoted by $\mathbf{A} = (A^1, A^2, ..., A^K) \in \mathcal{A} \subset \{0, 1\}^K$, where $A^k = 1$ indicates that the $k$th treatment is administered and $A^k = 0$ otherwise. Note that some combinations are infeasible to be considered in real applications, for example, many drug-drug interactions could lead to risks for patients outweighing the benefits \citep{rodrigues2019drug}. Therefore, we consider a subset $\mathcal{A}$ of all the possible $2^{K}$ combinations in the treatment rule. We may also denote the treatments by $\tilde{A} \in \{1, 2, ..., |\mathcal{A}|\}$ as categorical encodings, and use these two sets of notations interchangeably without ambiguity. The outcome of our interest is denoted by $Y \in \mathbb{R}$. Without loss of generality, we assume a larger value of $Y$ is preferable, for example, the shrinkage of tumor size. 

In causal inference, the potential outcome framework \citep{rubin1974estimating} is to describe the possible outcome after a certain treatment is assigned. We use $Y(\mathbf{A})$ to denote the potential outcome throughout the paper. Due to the ``fundamental problem of causal inference" \citep{holland1986statistics}, which indicates that only one potential outcome is observed for each subject, it is infeasible to estimate the subject-wise optimal treatment. Instead, our goal of estimating the optimal individualized treatment rule for combination treatments is to maximize the population-wise expected potential outcome, which is also known as the value function:
\begin{align} 
\label{value_func}
    \mathcal{V}(d) := \mathbb{E}[Y\{d(\mathbf{X})\}],
\end{align} 
where $d(\cdot): \mathcal{X}\rightarrow\mathcal{A}$ is an individualized treatment rule. The value function is defined as the expectation of the potential outcomes over the population distribution of $(\mathbf{X}, \mathbf{A}, Y)$ under $\mathbf{A} = d(\mathbf{X})$, which is estimable when the following causal assumptions \citep{rubin1974estimating} holds:
\begin{assumption}
\label{causal_assumption}
    (a) Stable Unit Treatment Value Assumption (SUTVA): $Y = Y(\mathbf{A})$;
    (b) No unmeasured confounders: $\mathbf{A} \indep Y(\mathbf{a}) | \mathbf{X}$, for any $\mathbf{a} \in \mathcal{A}$;
    (c) Positivity: $\mathbb{P}(\mathbf{A} = \mathbf{a}|\mathbf{X}) \ge p_{\mathcal{A}}$, $\forall \mathbf{a} \in \mathcal{A}, \forall \mathbf{X}\in\mathcal{X}$, for some $p_{\mathcal{A}} > 0$.
\end{assumption}

Assumption (a) is also referred to as ``consistency'' in causal inference, which assumes that the potential outcomes of each subject do not vary with treatments assigned to other subjects. The treatments are well-defined in that the same treatment leads to the same potential outcome. Assumption (b) states that all confounders are observed in pre-treatment covariates, so that the treatment and potential outcomes are conditionally independent given the pre-treatment covariates. Assumption (c) claims that for any pre-treatment covariates $\mathbf{X}$, each treatment can be assigned with a positive probability.

Based on these assumptions, the value function defined in (\ref{value_func}) can be identified as follows:
\begin{align}
\label{value_func_id}
    \mathcal{V}(d) = \mathbb{E}\{Y|\mathbf{A} = d(\mathbf{X})\} = \mathbb{E}\bigg\{\sum_{\mathbf{A}\in \mathcal{A}}\mathbb{E}(Y|\mathbf{X}, \mathbf{A})\mathbb{I}\{d(\mathbf{X}) = \mathbf{A}\}\bigg\},
\end{align}
where $\mathbb{I}(\cdot)$ is the indicator function. To maximize the value function, we can first estimate the conditional expectation $\mathbb{E}(Y|\mathbf{X} = \mathbf{x}, \mathbf{A} = \mathbf{a})$, namely the Q-function in the literature \citep{clifton2020q}. Then the optimal individualized treatment rule can be obtained by
\begin{align}
\label{decision_rule}
d^{*}(\mathbf{x}) \in \argmax_{\mathbf{a}\in \mathcal{A}}\mathbb{E}(Y|\mathbf{X} = \mathbf{x}, \mathbf{A} = \mathbf{a}).
\end{align}

From the perspective of the multi-arm treatments, the Q-function \citep{qian2011performance, qi2020multi, kosorok2019precision} can be formulated as:
\begin{align}
    \label{multiarm_problem}
    \mathbb{E}(Y|\mathbf{X}, \tilde{A}) = m(\mathbf{X}) + \sum_{l=1}^{|\mathcal{A}|}\delta_l(\mathbf{X})\mathbb{I}(\tilde{A}=l),
\end{align}
where $m(\mathbf{X})$ is the treatment-free effect representing a null effect without any treatment and functions $\delta_{l}(\mathbf{X})$'s are treatment effects for the $l$th treatment. {\large There are two major challenges when (\ref{multiarm_problem}) is applied to the combination treatments problem: First, if $\delta_{l}(\cdot)$'s are imposed to be some parametric model, for example, linear model \citep{qian2011performance, kosorok2019precision}, it could have severe misspecification issue, especially considering the complex nature of interaction effects of combination treatments. Second, as the number of treatments $K$ increases, the number of treatment-specific functions $\delta_{l}(\cdot)$'s could grow exponentially. Therefore, the estimation efficiency of the ITR based on Q-function (\ref{multiarm_problem}) could be severely compromised for either parametric or nonparametric models, especially in clinical trials or observational studies with limited sample sizes.}

In addition, considering the combination of multiple treatments expands the treatment space $\mathcal{A}$ and provides much more feasible treatment options. Therefore, each individual could have more choices rather than a yes-or-no as in the binary treatment scenario. Therefore, it is possible to consider accommodating realistic budget constraints while maintaining an effective outcome. In this paper, we further consider a population-level budget constraint as follows. Suppose costs over the $K$ treatments are $\mathbf{c} = (c_1, c_2, ..., c_{K})$, where $c_k$ denotes the cost for the $k$th treatment. Then the budget constraint for a population with a sample size $n$ is:
\begin{align}
    \label{budget_constraint}
    \mathcal{C}_{n}(d) := \frac{1}{n}\sum_{i=1}^{n}\mathbf{c}^Td(\mathbf{X}_i) \le B,
\end{align}
where $B$ is the average budget for each subject. This budget constraint is suitable for many policy-making problems such as welfare programs \citep{bhattacharya2012inferring} and vaccination distribution problem \citep{matrajt2021vaccine}.

\section{Methodology}
\label{sec: method}

In Section \ref{sec: dem}, we introduce the proposed Double Encoder Model (DEM) for estimating the optimal ITR for combination treatments. Section \ref{sec: bc-ITR} considers the optimal assignment of combination treatments under budget constraints. The estimation procedure and implementation details are provided in Section \ref{sec: est_imp}.

\subsection{Double Encoder Model for ITRs}
\label{sec: dem}

Our proposed Double Encoder Model (DEM) formulates the conditional expectation $\mathbb{E}(Y|\mathbf{X}, \mathbf{A})$, or the Q-function, as follows:
\begin{align}
    \label{model: dem}
    \mathbb{E}(Y|\mathbf{X}, \mathbf{A}) = m(\mathbf{X}) + \alpha(\mathbf{X})^T\beta(\mathbf{A}),
\end{align}
where $m(\cdot): \mathcal{X}\rightarrow \mathbb{R}$ is the treatment-free effects as in (\ref{multiarm_problem}), and $\alpha(\cdot): \mathcal{X} \rightarrow \mathbb{R}^{r}$ is an encoder that represents individuals' pre-treatment covariates in the $r$-dimensional latent space, which is called the covariate encoder. And $\beta(\cdot): \mathcal{A} \rightarrow \mathbb{R}^{r}$ is another encoder representing the combination treatment in the same $r$-dimensional latent space, named as the treatment encoder. In particular, these two encoders capture the unobserved intrinsic features of subjects and treatments; for instance, the covariates encoder $\alpha(\cdot)$ represents the patients' underlying health status, while the treatment encoder $\beta(\cdot)$ learns physiological mechanisms of the treatment. The inner product $\alpha(\mathbf{X})^T\beta(\mathbf{A})$ represents the concordance between subjects and treatments, hence representing the treatment effects on subjects.

From the perspective of function approximation, the covariates encoder $\alpha(\mathbf{X})$ learns the function bases of treatment effects, and the treatment encoder $\beta(\mathbf{A})$ learns the coefficients associated with those function bases. Consequently, the treatment effects are represented as the linear combinations of $r$ functions: 
\begin{align}
\delta_{l}(\mathbf{X}) = \sum_{i=1}^{r}\beta^{(i)}(\tilde{A}_l)\alpha^{(i)}(\mathbf{X}). \notag 
\end{align}
Note that the model for multi-arm treatments (\ref{multiarm_problem}) is a special case of the double encoder model (\ref{model: dem}) where $\alpha(\mathbf{X}) = (\delta_{1}(\mathbf{X}), ..., \delta_{|\mathcal{A}|}(\mathbf{X}))$ and $\beta(\tilde{A}) = (\mathbb{I}(\tilde{A} = \tilde{A}_1), ..., \mathbb{I}(\tilde{A} = \tilde{A}_{|\mathcal{A}|}))$ if $r = |\mathcal{A}|$. Another special case of (\ref{model: dem}) is the angle-based modeling \citep{zhang2020multicategory, qi2020multi, xue2021multicategory}, which has been applied to the estimation of the ITR for multi-arm treatments. In the angle-based framework, each treatment is encoded with a fixed vertex in the simplex, and each subject is projected in the latent space of the same dimension as the treatments so that the optimal treatment is determined by the angle between treatment vertices and the subject latent factors. However, the dimension of the simplex and latent space is $r = |\mathcal{A}| - 1$, which leads the angle-based modeling suffers from the same inefficiency issue as (\ref{multiarm_problem}).

Since different combination treatments could contain the same individual treatments, it is over-parameterized to model treatment effects for each combination treatment independently. For instance, the treatment effect of the combination of drug $A$ and drug $B$ is correlated with the individual treatment effects of drug $A$ and of drug $B$, respectively. Therefore, we seek to find a low-dimensional function space to incorporate the correlation of the combination treatments without over-parametrization. In the DEM (\ref{model: dem}), the dimension of the encoders output $r$ controls the complexity of the function space spanned by $\alpha^{(1)}(\cdot), ..., \alpha^{(r)}(\cdot)$. Empirically, the dimension $r$ is a tuning parameter, which can be determined via the hyper-parameter tuning procedure. In other words, the complexity of the DEM is determined by the data itself, rather than pre-specified. In addition, the reduced dimension also leads to a parsimonious model with fewer parameters, which permits an efficient estimation of treatment effects. Furthermore, we do not impose any parametric assumptions on $\alpha(\cdot)$, which allows us to employ flexible nonlinear or nonparametric models with $r$-dimensional output to avoid the potential misspecification of treatment effects.

Given the double encoder framework in (\ref{model: dem}), the treatment effects of the combination treatments share the same function bases $\alpha^{(1)}(\cdot)$, ..., $\alpha^{(r)}(\cdot)$. Therefore, the treatment encoder $\beta(\cdot)$ is necessary to represent all treatments in $\mathcal{A}$ so that $\alpha(\mathbf{X})^T\beta(\mathbf{A})$ can represent treatment effects for all treatments. Through this modeling strategy, we convert the complexity of $|\mathcal{A}|$ treatment-specific functions $\delta_{l}(\cdot)$'s in (\ref{multiarm_problem}) to the representation complexity of $\beta(\cdot)$ in that $\beta(\cdot)$ represents $|\mathcal{A}|$ treatments in $r$-dimensional latent space. As a result, we can reduce the complexity of the combination treatment problem and achieve an efficient estimation if an efficient representation (i.e. $r \ll |\mathcal{A}|$) of $|\mathcal{A}|$ treatments can be found.

In summary, the double encoder model (\ref{model: dem}) is a promising framework to tackle the two challenges in (\ref{multiarm_problem}) if covariates and treatment encoders can provide flexible and powerful representations of covariates and treatments, respectively, which will be elaborated in the following sections. Before we dive into the details of the covariates and treatment encoders, we first show the universal approximation property of the double encoder model, which guarantees its flexibility in approximating complex treatment effects.
\begin{theorem}
\label{thm: universal_approx}
    For any treatment effects $\delta_{l}(\mathbf{X}) \in \mathcal{H}^{2} = \{f: \int_{\mathbf{x}\in\mathcal{X}}|f^{(2)}(\mathbf{x})|^2d\mathbf{x} < \infty\}$, and for any $\epsilon > 0$, there exists $\alpha(\cdot): \mathcal{X}\rightarrow\mathbb{R}^{r}$ and $\beta(\cdot): \mathcal{X}\rightarrow\mathbb{R}^{r}$, where $K\le r\le |\mathcal{A}|$ such that
    \begin{align}
        \lVert\delta_{l}(\mathbf{X}) - \alpha(\mathbf{X})^T\beta(\tilde{A}_l)\rVert_{\mathcal{H}^2} \le \epsilon, \quad \text{for any } \tilde{A}_l \in \mathcal{A}. \notag
    \end{align}
\end{theorem}

The above theorem guarantees that the DEM (\ref{model: dem}) can represent the function space considered in (\ref{multiarm_problem}) sufficiently well given a sufficiently large $r$.

\subsubsection{Treatment Encoder}
\label{sec: trt_encoder}

In this section, we introduce the detailed modeling strategy for treatment encoder $\beta(\cdot)$. The treatment effects of combination treatments can be decoupled into two components: additive treatment effects, which is the sum of treatment effects from single treatments in combination; and interaction effects, which are the additional effects induced by the combinations of multiple treatments. Therefore, we formulate the treatment encoder as follows:
\begin{align}
\label{trt_encoder}
\begin{split}
    \beta(\mathbf{A}) &= \beta_{0}(\mathbf{A}) + \beta_{1}(\mathbf{A}) = \mathbf{W}\mathbf{A} + \beta_{1}(\mathbf{A}), \\
    \text{s.t.} & \quad\beta_1(\mathbf{A}) = 0, \quad \text{for any } \mathbf{A}\in\{\mathbf{A}: \sum_{k=1}^{K}A_k \ge 2\}, 
\end{split}    
\end{align}
where $\beta_0(\mathbf{A})$ and $\beta_1(\mathbf{A})$ are additive and interactive treatment encoders, respectively. In particular, $\beta_0(\mathbf{A})$ is a linear function with respect to $\mathbf{A}$, where $\mathbf{W} = (\mathbf{W}_{1}, \mathbf{W}_{2}, ..., \mathbf{W}_{K})\in \mathbb{R}^{r\times K}$ and $\mathbf{W}_{k}$ is the latent representation of the $k$th treatment. As a result, $\alpha(\mathbf{X})^T\beta_{0}(\mathbf{A}) = \sum_{k: \{A^{k} = 1\}}\mathbf{W}_k^T\alpha(\mathbf{X})$ are the additive treatment effects of the combination treatment $\mathbf{A}$. The constraints for $\beta_1(\cdot)$ ensures the identifiability of $\beta_0(\cdot)$ and $\beta_1(\cdot)$ such that any representation $\beta(\mathbf{A})$ can be uniquely decoupled into $\beta_0(\mathbf{A})$ and $\beta_1(\mathbf{A})$.

The interaction effects are challenging to estimate in combination treatments. A naive solution is to assume that interaction effects are ignorable, which leads the additive treatment encoder $\beta_{0}(\mathbf{A})$ to be saturated in estimating the treatment effects of combination treatments. However, interaction effects are widely perceived in many fields such as medicine \citep{stader2020stopping, li2018assessment}, psychology \citep{caspi2010genetic}, and public health \citep{braveman2011broadening}. Statistically, ignoring the interaction effects could lead to inconsistent estimation of the treatment effects \citep{zhao2023covariate} and the ITR \citep{liang2018estimating}. Hence, it is critical to incorporate the interaction effects in estimating the ITR for combination treatments.

A straightforward approach to model the interactive treatment encoder $\beta_1(\mathbf{A})$ is similar to the additive treatment encoder $\beta_0(\mathbf{A})$, which we name as the treatment dictionary. Specifically, a matrix $\mathbf{V} = (\mathbf{V}_1, \mathbf{V}_2, ..., \mathbf{V}_{|\mathcal{A}|})\in\mathbb{R}^{r\times |\mathcal{A}|}$ is a dictionary that stores the latent representations of each combination treatment so that $\beta_1(\mathbf{A})$ is defined as follows
\begin{align}
\label{trt_dict}
\beta_{0}(\mathbf{A}) = \mathbf{V}\mathbf{e}_{\tilde{A}},
\end{align}
where $\mathbf{e}_{\tilde{A}}$ is the one-hot encoding of the categorical representation of $\mathbf{A}$. Since the number of possible combination treatments $|\mathcal{A}|$ could grow exponentially as $K$ increases, the parameters of $\mathbf{V}$ could also explode. Even worse, each column $\mathbf{V}_{l}$ can be updated only if the associated treatment $\tilde{A}_{l}$ is observed. Given a limited sample size, each treatment could be only observed a few times in the combination treatment scenarios, which leads the estimation efficiency of $\mathbf{V}$ to be severely compromised. The same puzzle is also observed in other methods. For the Q-function in(\ref{multiarm_problem}), the parameters in $\delta_{l}(\mathbf{X})$ can be updated only if $\tilde{A}_{l}$ is observed; In the Treatment-Agnostic Representation Network (TARNet) \citep{shalit2017estimating} and the Dragonnet \citep{shi2019adapting}, each treatment is associated with an independent set of regression layers to estimate the treatment-specific treatment effects, which results in inefficiency estimation for combination treatment problems.

\begin{figure}
    \centering
    \includegraphics[width=0.9\textwidth]{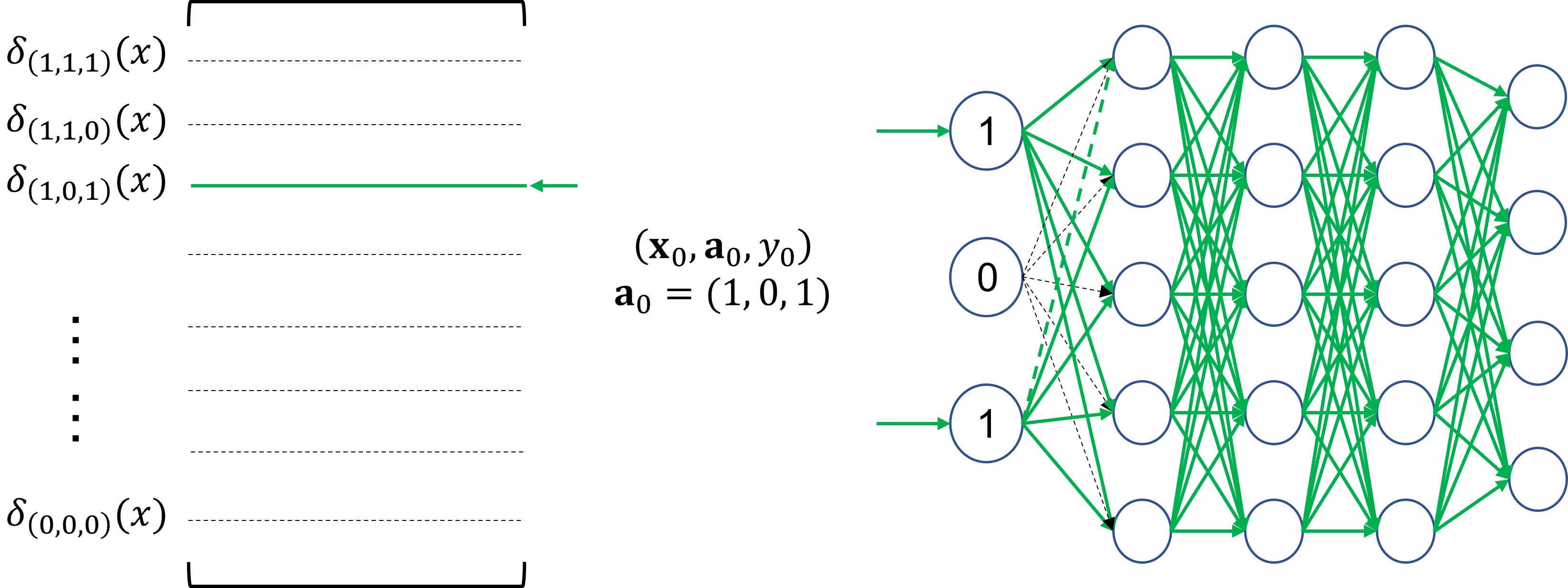}
    \caption{The left panel shows the parameter update scheme in the additive model (\ref{multiarm_problem}), the treatment dictionary (\ref{trt_dict}), TARNet \citep{shalit2017estimating} and dragonnet \citep{shi2019adapting}. Only the treatment-specific parameters corresponding to $\mathbf{a}_0$ are updated. The right panel shows the parameter-sharing feature of the neural network interactive treatment encoder (\ref{trt_encoder_nn}). All parameters except for non-activated input parameters are updated based on the gradient with respect to the observation $(\mathbf{x}_0, \mathbf{a}_0, y_{0})$.}
    \label{fig: parameter_sharing}
\end{figure}

In order to overcome the above issue, we propose to utilize the feed-forward neural network \citep{goodfellow2016deep} to learn efficient latent representations in the $r$-dimensional space. Specifically, the interactive treatment encoder is defined as
\begin{align}
\label{trt_encoder_nn}
    \beta_{1}(\mathbf{A}) = \mathcal{U}_{L} \circ \sigma \circ ... \circ \sigma \circ \mathcal{U}_{1}(\mathbf{A}),
\end{align}
where $\mathcal{U}_{l}(\mathbf{x}) = \mathbf{U}_{l}\mathbf{x} + \mathbf{b}_l$ is the linear operator with the weight matrix $\mathbf{U}_l \in \mathbb{R}^{r_{l}\times r_{l-1}}$ and the biases $\mathbf{b}_l$. The activation function is chosen as ReLU function $\sigma(\mathbf{x}) = \max(\mathbf{x}, 0)$ in this paper. An illustration of the neural network interactive treatment encoder is shown in Figure \ref{fig: parameter_sharing}. Note that all parameters in (\ref{trt_encoder_nn}) are shared among all possible treatments, so all of the weight matrices and biases in (\ref{trt_encoder_nn}) are updated regardless of the input treatment, which could improve the estimation efficiency, even though (\ref{trt_encoder_nn}) may include more parameters than the treatment dictionary (\ref{trt_dict}). As a result, the double encoder model with (\ref{trt_encoder_nn}) not only guarantees a faster convergence rate (with respect to $K$) of the value function but also improves the empirical performance especially when $K$ is large or sample size $n$ is small, which will be shown in numerical studies and real data analysis. A direct comparison of the neural network interactive treatment encoder (\ref{trt_encoder_nn}) and the treatment encoder (\ref{trt_dict}), the additive model (\ref{multiarm_problem}), TARNet \citep{shalit2017estimating} and Dragonnet \citep{shi2019adapting} are also shown in Figure \ref{fig: parameter_sharing}.

Although the interactive treatment encoder (\ref{trt_encoder_nn}) allows an efficient estimation, it is not guaranteed to represent up to $|\mathcal{A}|$ interaction effects. In the treatment dictionary (\ref{trt_dict}), columns $\mathbf{V}_{l}$'s are free parameters to represent $|\mathcal{A}|$ treatments without any constraints. However, an ``under-parameterized'' neural network is not capable of representing $|\mathcal{A}|$ treatments in $r$-dimensional space. For example, if there are three treatments to be combined ($K=3$), and the treatment effects are sufficiently captured by one-dimensional $\alpha(\mathbf{X})$ with different coefficients ($r = 1$). We use the following one-hidden layer neural network to represent the treatment in $\mathbb{R}$:
\begin{align}
\label{toy_nn}
\beta_{1}(\mathbf{A}) = u_{2}\sigma(\mathbf{U}_{1}\mathbf{A} + b_1) + b_2, 
\end{align}
where $u_2, b_2, b_1 \in \mathbb{R}$ are scalars and $\mathbf{U}_{1}\in \mathbb{R}^{1\times 3}$. In other words, the hidden layer only includes one node. In the following, we show that this neural network can only represent restricted interaction effects:
\begin{prop}
\label{prop: toy_example}
The one-hidden layer neural network (\ref{toy_nn}) can only represent the following interaction effects: (a) $\beta_1(\mathbf{A}) \ge 0$ or $\beta_1(\mathbf{A}) \le 0$ for all $\mathbf{A} \in \mathcal{A}$; (b) $\beta_1(\mathbf{A})$ takes the same values for all combinations of two treatments.
\end{prop}

\textcolor{black}{The proof of Proposition \ref{prop: toy_example} is provided in the supplementary materials. Based on the above observation, it is critical to guarantee the representation power of $\beta_1(\mathbf{A})$ to incorporate flexible interaction effects. In the following, we establish a theoretical guarantee of the representation power of $\beta_1(\cdot)$ under a mild assumption on the widths of neural networks:
\begin{theorem}
\label{thm: rep_power}
For any treatment $\mathbf{A} \in \mathcal{A} \subset \{0, 1\}^{K}$, if $\beta_1(\cdot)$ is a 3-layer fully-connected neural network defined in (\ref{trt_encoder_nn}) satisfying $4[r_{1}/4][r_{2}/4r] \ge |\mathcal{A}|$, then there exist parameters $\{\mathbf{U}_{l}, \mathbf{b}_{l}, l=1, 2, 3\}$, such that $\beta(\mathbf{A})$ satisfies the identifiability constraints and can take any values in $\mathbb{R}^{r}$. 
\end{theorem}}

\textcolor{black}{The above result is adapted from the recent work on the memorization capacity of neural networks \citep{yun2019small, bubeck2020network}. Theorem \ref{thm: rep_power} shows that if there are $\Omega(2^{K/2}r^{1/2})$ hidden nodes in neural networks, then it is sufficient to represent all possible interaction effects in $\mathbb{R}^{r}$. However, obtaining the parameter set $\{\mathbf{U}_{l}, \mathbf{b}_{l}, l=1, 2, 3\}$ in Theorem \ref{thm: rep_power} via the optimization algorithm is not guaranteed due to the non-convex loss surface of the neural networks. In practice, the neural network widths in Theorem \ref{thm: rep_power} can be a guide, and choosing a wider network is recommended to achieve better empirical performance.}

In summary, we propose to formulate the treatment encoder as two decoupled parts: the additive treatment encoder and the interactive encoder. We provide two options for the interactive treatment encoder: the treatment dictionary and the neural network, where the neural network can improve the asymptotic convergence rate and empirical performance with guaranteed representation power. In the numerical studies, we use the neural network interactive treatment encoder for our proposed method, and a comprehensive comparison between the treatment dictionary and the neural network is provided in the supplementary materials.

\subsubsection{Covariates Encoder}
\label{sec: cov_encoder}

As we introduced in (\ref{model: dem}), the covariates encoder $\alpha(\cdot): \mathcal{X}\rightarrow \mathbb{R}^{r}$ constitutes the function bases of the treatment effects for all combination treatments. In other words, the treatment effects represented in (\ref{model: dem}) lie in the space spanned by $\alpha^{(1)}(\mathbf{X}), ..., \alpha^{(r)}(\mathbf{X})$. Therefore, it is critical to consider a sufficiently large and flexible function space to accommodate the highly complex treatment effects and avoid possible model misspecification. In particular, we adopt three nonlinear or nonparametric models for covariates encoders: polynomial, B-Spline \citep{hastie2009elements}, and neural network \citep{goodfellow2016deep}.

First of all, we introduce the $\alpha(\mathbf{X})$ as a feed-forward neural network defined as follows:
\begin{align}
\label{cov_encoder}
    \alpha(\mathbf{X}) = \mathcal{T}_{L}\circ \sigma \circ ... \circ \sigma \circ \mathcal{T}_{1}(\mathbf{X}),
\end{align}
$\mathcal{T}_{l}(\mathbf{x}) = \mathbf{T}_{l}\mathbf{x} + \mathbf{c}_l$ is the linear operator with the weight matrix $\mathbf{T}_l \in \mathbb{R}^{r_{l}\times r_{l-1}}$ and the biases $\mathbf{c}_l$. The activation function is chosen as ReLU function $\sigma(\mathbf{x}) = \max(\mathbf{x}, 0)$ in this paper. Note that the depth and the width of the covariates encoder $\alpha(\cdot)$ are not necessarily identical to those of the interactive treatment encoder $\beta_{1}(\cdot)$, and these are all tuning parameters to be determined through hyper-parameter tuning.

\begin{figure}
    \centering
    \includegraphics[width=0.3\textwidth]{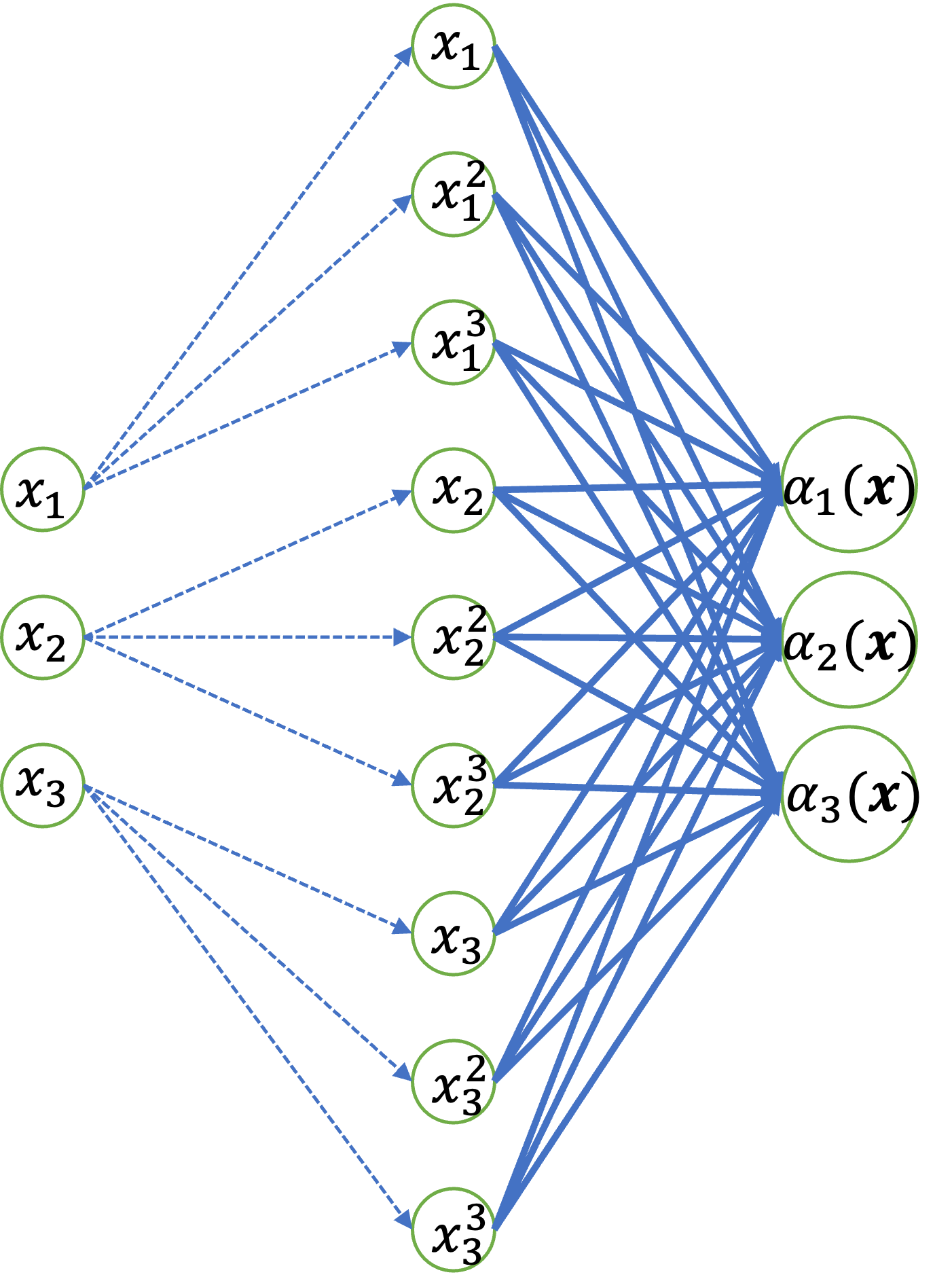}
    \caption{Model structure of the polynomial covariate encoder. Dashed lines indicate the fixed polynomial expansion procedures, and solid lines are trainable parameters for the linear combination of polynomials.}
    \label{fig: poly_cov_encoder}
\end{figure}

Even though neural networks achieve superior performance in many fields, their performance in small sample size problems, such as clinical trials or observational studies in medical research, is still deficient. In addition, neural networks lack interpretability due to the nature of their recursive composition; therefore, the adoption of neural networks in medical research is still under review. Here, we propose the polynomial and B-Spline covariates encoders to incorporate nonlinear treatment effects for better interpretation. For the polynomial covariates encoder, we first expand each covariate $x_i$ into a specific order of polynomials $(x_i, x_i^2, ..., x_i^d)$ where $d$ is a tuning parameter. Then we take the linear combinations of all polynomials as the output of the covariate encoders. Figure \ref{fig: poly_cov_encoder} provides an example of the polynomial covariate encoder with $d=3$. Similarly, as for the B-spline covariates encoder, we first expand each covariate into B-spline bases, where the number of knots and the spline degree are tuning parameters. Likewise, linear combinations of these B-spline bases are adopted as the output of the encoder. Although both polynomial and B-spline covariate encoders can accommodate interaction terms among the polynomial bases or B-spline bases for a better approximation for multivariate functions, exponentially increasing parameters need to be estimated as the dimension of covariates or the degree of bases increases. In the interest of computation feasibility, we do not consider interaction terms in the following discussion.

\subsection{Budget Constrained Individualized Treatment Rule}
\label{sec: bc-ITR}

In this section, we consider to optimize the assignment of combination treatments under the budget constraints, where the total cost constraints are imposed on a population with a sample size $n$.

We first introduce budget constrained ITR for binary treatments. Suppose we have a treatment ($A = 1$) and a control ($A = -1$), and there are only $b\%$ subjects which can be treated with the treatment $A = 1$. \citet{luedtke2016optimal} define a contrast function $\eta(\mathbf{x}) = \delta(\mathbf{x}, 1) - \delta(\mathbf{x}, -1)$, and the corresponding ITR under the budget constraint is $d(\mathbf{x}, b) = I(\eta(\mathbf{x}) \ge q_b)$, where $q_b$ is the $b\%$ quantile of the distribution of $\eta(\mathbf{x})$ for a given population with a finite sample size.

Estimating the optimal individualized treatment rule for combination treatments under budget constraints is challenging. First of all, the contrast function is no longer a valid tool to measure the treatment importance to each subject. Given the exponentially increasing choices of combination treatments, the number of contrast functions increases exponentially and the pairwise comparisons do not suffice to determine the optimal assignment. Second, costs over different channels may differ significantly, which makes the quantile no longer an effective criterion for allocating budgets.

In the following, we consider the constrained ITR problem for a finite population:
\begin{equation}
\begin{aligned}
\label{constrained_problem}
    \max_{d} \hspace{2mm} \mathcal{V}_n(d) \hspace{2mm}
    s.t. \hspace{2mm} \mathcal{C}_{n}(d)\le B,
\end{aligned}
\end{equation}
where $\mathcal{V}_{n}$ and $\mathcal{C}_n$ are defined on a pre-specified population with a sample size $n$. Here, covariates $\mathbf{x}_i$ ($i=1, 2, ..., n$) are treated as fixed covariates. Based on model formulation (\ref{model: dem}), maximizing the objective function of (\ref{constrained_problem}) is equivalent to:
\begin{align}
    \argmax_{d}\mathcal{V}_{n}(d) &= \argmax_{d}\frac{1}{n}\sum_{i=1}^{n}[m(\mathbf{x}_i) + \alpha(\mathbf{x}_i)^T\beta(d(\mathbf{x}_i))] \notag \\
    &= \argmax_{d}\frac{1}{n}\sum_{i=1}^{n}\alpha(\mathbf{x}_i)^T\beta(d(\mathbf{x}_i)) \notag \\
    &= \argmax_{\{d_{ij}\}}\frac{1}{n}\sum_{i=1}^{n}\sum_{j=1}^{|\mathcal{A}|}\delta_{ij}d_{ij}, \notag
\end{align}
 where $\delta_{ij} = \alpha(\mathbf{x}_i)^T\beta(\mathbf{a}_j)$ denotes the treatment effects of the $j$th combination treatment on the $i$th subject, and $d_{ij} = I\{d(\mathbf{x}_i)=\mathbf{a}_j\} \in \{0, 1\}$ indicates whether the $i$th subject receives the $j$th combination treatment. Since one subject can only receive one combination treatment, we impose the constraint to be $\sum_{j}d_{ij} = 1$. Similarly, budget constraints can be formulated as $\frac{1}{n}\sum_{i=1}^{n}\sum_{j=1}^{|\mathcal{A}|} c_{\tilde{a}_j}d_{ij} \le B$, where $c_{\tilde{a}_j}$ is the cost of treatment $\tilde{a}_j$ calculated from the cost vector $\mathbf{c}$. The constrained individualized treatment rule can be solved as follows:

\begin{equation}
\begin{aligned}
\label{empirical_constrained_problem}
    \max_{\{d_{ij}\}}& \hspace{2mm}\frac{1}{n}\sum_{i=1}^{n}\sum_{j=1}^{|\mathcal{A}|}\delta_{ij}d_{ij}, \\
    s.t. \hspace{2mm} \frac{1}{n}\sum_{i=1}^{n}\sum_{j=1}^{|\mathcal{A}|} c_{\tilde{a}_j}d_{ij} \le B, 
   &\hspace{2mm}\sum_{j}d_{ij} = 1, \hspace{2mm}
   \hspace{2mm} d_{ij} \in \{0, 1\}, \hspace{2mm} \text{ for any } i,j.
\end{aligned}
\end{equation}
The above optimization problem is equivalent to a multi-choice knapsack problem \citep{kellerer2004multidimensional}. For a binary treatment setting, the solution of (\ref{empirical_constrained_problem}) is the quantile of $\eta(\mathbf{X})$, which is a special case of our formulation.

To understand the connection between constrained individualized treatment rule and the multi-choice knapsack problem, we notice that the priority of the treatment is associated with the definition of dominance in the multi-choice knapsack problem: for any $i \in \{1, 2, ..., n\}$, if $\delta_{ik} > \delta_{il}$ and $c_{\mathbf{a}_k} < c_{\mathbf{a}_l}$, the treatment $l$ is dominated by the treatment $k$. In other words, the treatment $k$ achieves a better outcome than the treatment $l$ with a lower cost. Thus, the dominance property could be an alternative to the contrast functions in combination treatment settings.

Here $\delta_{ij}$ indicates the treatment effects, and the parametric assumptions are not required, so this framework is also applicable for other methods providing estimations of treatment effects such as the $l_1$-penalized least-square \citep{qian2011performance}, and the outcome weighted learning with multinomial deviance \citep{huang2019multicategory}. However, the objective function in (\ref{constrained_problem}) depends on the estimation of $\delta_{ij}$, and we show that the value reduction under budget constraints is bounded by the estimation error of $\delta_{ij}$'s in Theorem \ref{thm: value_reduction_constraint}. Consequently, estimation bias in $\delta_{ij}$ could lead to biased results in (\ref{empirical_constrained_problem}). Since the proposed model (\ref{model: dem}) provides an efficient and accurate estimation of treatment effects, it also results in a more favorable property for solving the budget constrained individualized treatment rule for combination treatments. 

\subsection{Estimation and Implementation}
\label{sec: est_imp}

In this section, we introduce the estimation and hyper-parameter tuning procedures of the proposed method for unconstrained and constrained individualized treatment rule for combination treatment.

\subsubsection{Estimation of the Double Encoder Model}
First of all, we propose the following doubly robust estimator for the treatment effects: 
\begin{align}
\label{loss_func}
    \hat{\alpha}(\cdot), \hat{\beta}(\cdot) = \argmin_{\alpha(\cdot), \beta(\cdot)}\mathbb{E}\bigg\{\frac{1}{\hat{\mathbb{P}}(\mathbf{A}_i|\mathbf{X_i})}(Y_i - \hat{m}(\mathbf{X}_i) - \alpha(\mathbf{X}_i)^T\beta(\mathbf{A}_i))^2\bigg\}, 
\end{align}
where $\hat{\mathbb{P}}(\mathbf{A}_{i}|\mathbf{X_i})$ is a working model of the propensity score specifying the probability of treatment assignment given pre-treatment covariates, and $\hat{m}(\mathbf{X}_i)$ is a working model of treatment-free effects. The inverse probability weights given by the propensity scores balance the samples assigned to different combination treatments, assumed to be equal under the randomized clinical trial setting. By removing the treatment-free effects $m(\mathbf{x})$ from the responses before we estimate the treatment effects, the numerical stability can be improved and the estimator variance is reduced. This is also observed in \citep{zhou2017residual, fu2016estimating}. Furthermore, the estimator in (\ref{loss_func}) is doubly-robust in that if either $\hat{\mathbb{P}}(\cdot|\cdot)$ or $\hat{m}(\cdot)$ is correctly-specified, $\hat{\alpha}(\cdot)^T\hat{\beta}(\cdot)$ is a consistent estimator of the treatment effects, and a detailed proof is provided in the supplemental material. This result extends the results in \citep{meng2020doubly} from binary and multiple treatments to combination treatments. Empirically, we minimize the sample average of the loss function (\ref{loss_func}) with additional penalties: for the additive treatment encoder, $L_2$ penalty is imposed to avoid overfitting; for the interactive treatment encoder, $L_1$ penalty is added since the interaction effects are usually sparse \citep{wu2011experiments}.

In this work, the working model of the propensity score is obtained via the penalized multinomial logistic regression \citep{friedman2010regularization} as a working model. Specifically, the multinomial logistic model is parameterized by $\gamma_1, \gamma_2, ..., \gamma_{2^{K}} \in \mathbb{R}^{p}$:
\begin{align}
    \mathbb{P}(\tilde{\mathbf{A}} = k|\mathbf{x}) = \frac{\exp(\gamma_{k}^T\mathbf{x})}{\sum_{k'=1}^{2^{K}}\exp(\gamma_{k'}^T\mathbf{x})}. \notag
\end{align}
The parameters $\gamma_{k}$'s can be estimated by maximizing the likelihood:
\begin{align}
    \max_{\gamma_{1},...,\gamma_{2^{K}}}\frac{1}{n}\sum_{i=1}^{n}\bigg[\sum_{k=1}^{2^{K}}\gamma_{k}^T\mathbf{x}_{i}I(\tilde{\mathbf{A}}=k) - \log\{\sum_{k'=1}^{2^{K}}\exp(\gamma_{k'}^T\mathbf{x_i})\}\bigg] - \lambda\sum_{j=1}^{p}(\sum_{k=1}^{2^{K}}\gamma_{kj}^{2})^{1/2}, \notag
\end{align}
where the group Lasso \citep{meier2008group} is used to penalize parameters across all treatment groups. A potential issue of propensity score estimation is that the estimated probability could be negligible when there are many possible treatments, which leads to unstable estimators for treatment effects. To alleviate the limitation on inverse probability weighting, we stabilize the propensity scores \citep{xu2010use} by multiplying the frequency of the corresponding treatment to the weights.

For the estimation of the treatment-free effects $m(\cdot)$, we adopt a two-layer neural network:
\begin{align}
\label{treatment_free_function}
    m(\mathbf{x}) = (\mathbf{w}_{m}^{2})^{T}\sigma(\mathbf{W}_{m}^{1}\mathbf{x}), 
\end{align}
where $\mathbf{w}^2_{m} \in \mathbb{R}^{h}$ and $\mathbf{W}_{m}^{1} \in \mathbb{R}^{h\times p}$ are weight matrices, and $\sigma(x)$ is the ReLu function. The width $h$ controls the complexity of this model. The weight matrices are estimated through minimizing:
\begin{align}
    \min_{\mathbf{w}_{m}^{2}, \mathbf{W}_{m}^{1}} \frac{1}{n}\sum_{i=1}^{n}[y_i - (\mathbf{w}_{m}^{2})^{T}\sigma(\mathbf{W}_{m}^{1}\mathbf{x})]^{2}. \notag
\end{align}

Given working models $\hat{m}(\mathbf{x})$ and $\hat{\mathbb{P}}(\mathbf{a}|\mathbf{x})$ for treatment-free effects and propensity scores, we propose to optimize the double encoder alternatively. The detailed algorithm is listed in Algorithm \ref{double_encoder_alg}. Specifically, we employ the Adam optimizer \citep{kingma2014adam} for optimizing each encoder: covariate encoder $\alpha(\mathbf{x})$, additive treatment encoder $\beta_{1}(\mathbf{a})$, and non-parametric treatment encoder $\beta_{2}(\mathbf{a})$. To stabilize the optimization during the iterations, we utilize the exponential scheduler \citep{patterson2017deep} which decays the learning rate by a constant per epoch. In all of our numerical studies, we use 0.95 as a decaying constant for the exponential scheduler. To ensure the identifiability of treatment effects, we also require a constraint on the treatment encoder $\beta(\cdot)$ such that $\sum_{\mathbf{a}\in\mathcal{A}}\beta(\mathbf{a}) = 0$. To satisfy this constraint, we add an additional normalization layer before the output of $\beta(\cdot)$. The normalization layer subtracts the weighted mean vector where the weight is given by the reciprocal of the combination treatment occurrence in the batch. Since this operation only centers the outputs, the theoretical guarantee for $\beta(\cdot)$ in Section \ref{sec: theory} still holds. Once our algorithm converges, we obtain the estimation for $\alpha(\cdot)$ and $\beta(\cdot)$, and also the estimated individualized treatment rule $\hat{d}(\cdot)$ by (\ref{decision_rule}).

\begin{algorithm}
    \caption{Double encoder model training algorithm}
    \label{double_encoder_alg}
    \begin{algorithmic}
        \State \textbf{Input}: Training dataset $(\mathbf{x}_i, \mathbf{a}_i, \mathbf{y}_i)_{i=1}^{n}$, working models $\hat{m}(\mathbf{x}), \hat{\mathbb{P}}(\mathbf{a}|\mathbf{x})$, hyper-parameters including network structure-related hyper-parameters (e.g., network depth $L_{\alpha}, L_{\beta}$, network width $r_{\alpha}, r_{\beta}$, encoder output dimension $r$), optimization-related hyper-parameters (e.g., additive treatment encoder penalty coefficients $\lambda_{a}$, interactive treatment encoder penalty coefficients $\lambda_{i}$, mini-batch size $B$, learning rate $\eta$, and training epochs $E$).
        \State \textbf{Initialization}: Initialize parameters in $\hat{\alpha}^{(0)}(\mathbf{x})$, $\hat{\beta}^{(0)}_0(\mathbf{a})$, and $\hat{\beta}^{(0)}_1(\mathbf{a})$.
        \State \textbf{Training}: 
        \For{e in 1: E}
            \For{Mini-batch sampled from $(\mathbf{x}_i, \mathbf{a}_i, \mathbf{y}_i)_{i=1}^{n}$}
            \State $\hat{\alpha}^{(e)}(\mathbf{x}) = \argmin\frac{1}{B}\sum\frac{1}{\hat{\mathbb{P}}(\mathbf{a}_i|\mathbf{x_i})}\bigg\{y_i - \hat{m}(\mathbf{x}_i) - \alpha(\mathbf{x}_i)^T(\hat{\beta}_0^{(e-1)}(\mathbf{a}_i) + \hat{\beta}_1^{(e-1)}(\mathbf{a}_i))\bigg\}^2$
            \State $\hat{\beta}_{1}^{(e)}(\mathbf{a}) = \argmin\frac{1}{B}\sum\frac{1}{\hat{\mathbb{P}}(\mathbf{a}_i|\mathbf{x_i})}\bigg\{y_i - \hat{m}(\mathbf{x}_i) - \hat{\alpha}^{(e)}(\mathbf{x}_i)^T(\beta_0(\mathbf{a}_i) + \hat{\beta}_1^{(e-1)}(\mathbf{a}_i))\bigg\}^2 + \lambda_{a}\lVert \beta_0 \rVert_2$
            \State $\hat{\beta}_{2}^{(e)}(\mathbf{a}) = \argmin\frac{1}{B}\sum\frac{1}{\hat{\mathbb{P}}(\mathbf{a}_i|\mathbf{x_i})}\bigg\{y_i - \hat{m}(\mathbf{x}_i) - \hat{\alpha}^{(e)}(\mathbf{x}_i)^T(\hat{\beta}_0^{(e)}(\mathbf{a}_i) + \beta_1(\mathbf{a}_i))\bigg\}^2 + \lambda_{i}\rVert \beta_1\rVert_1$
            \EndFor
        \EndFor
    \end{algorithmic}
\end{algorithm}

In addition, successful neural network training usually requires careful hyper-parameter tuning. The proposed double encoder model includes multiple hyper-parameters: network structure-related hyper-parameters (e.g., network depth $L_{\alpha}$ and $L_{\beta}$, network width $r_{\alpha}$ and $r_{\beta}$, encoder output dimension $r$), optimization-related hyper-parameters (e.g., additive treatment encoder penalty coefficients $\lambda_{a}$, interactive treatment encoder penalty coefficients $\lambda_{i}$, mini-batch size $B$, learning rate $\eta$, and training epochs $E$). These hyper-parameters induce an extremely large search space, which makes the grid search method \citep{yu2020hyper} practically infeasible. Instead, we randomly sample 50 hyper-parameter settings in each experiment over the pre-specified search space (detailed specification of hyper-parameter space is provided in the supplementary materials), and the best hyper-parameter setting is selected if it attains the largest value function on an independent validation set. Furthermore, due to the non-convexity of the loss function, the convergence of the algorithm also relies heavily on the parameter initialization. In the supplementary materials, we provide detailed analyses for numerical results under different parameter initializations.

\subsubsection{Budget-constrained ITR estimation}

In the following, we introduce our procedure for the budget-constrained individualized treatment rule for combination treatment (\ref{constrained_problem}) estimation. We use the plug-in estimates $\hat{\alpha}(\mathbf{x}_i)^T\hat{\beta}(\mathbf{a}_j)$ for $\delta_{ij}$, and calculate the cost for each combination treatment from the cost vector $\mathbf{c}$ by $c_{\mathbf{a}_j} = \mathbf{a}_{j}^T\mathbf{c}$. Then we apply the dynamic programming algorithm to solve (\ref{constrained_problem}) with plug-in $\delta_{ij}$'s. Although the multi-choice knapsack problem is a NP-hard problem, we can still solve it within pseudo-polynomial time \citep{kellerer2004multidimensional}. Specifically, we denote $\hat{Z}_l(b)$ as the optimal value $\frac{1}{n}\sum_{i=1}^{l}\sum_{j=1}^{|\mathcal{A}|}\hat{\delta}_{ij}d_{ij}$ for the first $l$ subjects with budget constraints $\frac{1}{n}\sum_{i=1}^{l}\sum_{j=1}^{|\mathcal{A}|}c_{\tilde{a}_j}d_{ij} \le b$. Let $\hat{Z}_l(b) = -\infty$ if no solution exists and $\hat{Z}_0(b) = 0$. We define the budget space as $\mathcal{B} = \{b: 0 \le b \le B\}$ including all possible average costs for $n$ subjects, where $0$ is the minimal cost if no treatment is applied to subjects, and the maximal cost is our specified budget $B$. Once the iterative algorithm ends, the optimal objective function is obtained as $\hat{Z}_{n}(B)$ and the optimal treatment assignment is the output $\{d_{ij}: i=1,...,n, j=1,...,|\mathcal{A}|\}$. The detailed algorithm is illustrated in Algorithm \ref{dp_for_constraints}.

\begin{algorithm}
\caption{Pseudo code of dynamic programming algorithm}
\label{dp_for_constraints}
\begin{algorithmic}[1]
\State Input: Treatment effects $\{\hat{\delta}_{ij}: i=1,2,...n, j=1,...,|\mathcal{A}|\}$, cost $\{c_{\tilde{A}_j}: j=1,...,|\mathcal{A}|\}$, budget $B$.
\State Initialize: $\hat{Z}_0(b) \gets 0$, \text{ for } $b \in \mathcal{B} =  \{b: 0 \le b \le B\}$
\While{$l < n$}
\State $l \gets l + 1$
\For{$b \in \mathcal{B}$}
\State $\hat{Z}_{l}(b) \gets \max_{j: b > c_{\tilde{A}_j}}\hat{Z}_{l-1}(b-c_{\tilde{A}_{j}}) + \hat{\delta}_{lj}/n$
\State $d_{lj} \gets 1$ if $j = \argmax_{j: b > c_{\tilde{A}_j}}\hat{Z}_{l-1}(b-c_{\tilde{A}_{j}}) + \hat{\delta}_{lj}/n$; Otherwise, $d_{lj} \gets 0$
\EndFor
\EndWhile
\State Output: $\{d_{ij}: i=1,...,n, j=1,...,|\mathcal{A}|\}$
\end{algorithmic}
\end{algorithm}

\section{Theoretical Guarantees}
\label{sec: theory}
In this section, we establish the theoretical properties of the ITR estimation for combination treatments and the proposed method. First, we establish the value reduction bound for the combination treatments, either with or without budget constraints. Second, we provide a non-asymptotic excess risk bound for the double encoder model, which achieves a faster convergence rate compared with existing methods for multi-arm treatment problems.

\subsection{Value Reduction Bound}

\color{black}
The value reduction is the difference between the value functions of the optimal individualized treatment rule and of the estimated individualized treatment rule. The value function under a desirable ITR is expected to converge to the value function under the optimal ITR when the sample size goes to infinity. Prior to presenting the main results, we introduce some necessary notations. The conditional expectation of the outcome $Y$ given the subject variable $\mathbf{X}$ and the treatment $\mathbf{A}$ is denoted by $Q(\mathbf{X}, \mathbf{A}) = \mathbb{E}[Y|\mathbf{X}, \mathbf{A}]$, and the treatment-free effects can be rewritten as $m(\mathbf{X}) = \mathbb{E}[Q(\mathbf{X}, \mathbf{A})|\mathbf{X}]$, and the treatment effects can be denoted as $\delta(\mathbf{X}, \mathbf{A}) = Q(\mathbf{X}, \mathbf{A}) - m(\mathbf{X})$. In particular, the true and the estimated treatment effects are denoted as $\delta^*(\cdot, \cdot)$ and $\hat{\delta}(\cdot, \cdot)$, respectively. In addition, we introduce an assumption on the treatment effects:
\begin{assumption}
    \label{margin_assumption}
    For any $\epsilon > 0$, there exist some constant $C > 0$ and $\gamma > 0$ such that 
    \begin{align}
    \label{margin_condition}
        \mathbb{P}(\max_{\mathbf{A}, \mathbf{A}'\in\mathcal{A}}|\delta^{*}(\mathbf{X}, \mathbf{A}) - \delta^{*}(\mathbf{X}, \mathbf{A}')|\le \epsilon) \le C\epsilon^{\gamma}.
    \end{align}
\end{assumption}

Assumption \ref{margin_assumption} is a margin condition characterizing the behavior of the boundary between different combination treatments. A larger value of $\gamma$ indicates that the treatment effects are differentiable with a higher probability, suggesting it is easier to find the optimal individualized treatment rule. Similar assumptions are also required in the literature \citep{qian2011performance, zhao2012estimating, qi2020multi} to achieve a faster convergence rate of the value reduction bound. 

The following theorem shows that the value reduction is bounded by the estimation error of the treatment effects, and the convergence rate can be improved if Assumption \ref{margin_assumption} holds:

\begin{theorem}
\label{theorem1}
    Suppose the treatment effects $\delta^{*}(\cdot, \cdot) \in \mathcal{H}^2$. For any estimator $\hat{\delta}(\cdot, \cdot)$, and the corresponding decision rule $\hat{d}$ such that $\hat{d}(\mathbf{X}) \in \argmax_{\mathbf{A}\in \mathcal{A}}\hat{\delta}(\mathbf{X}, \mathbf{A})$, we have
    \begin{align}
    \label{theorem1_bound1}
        \mathcal{V}(d^{*}) - \mathcal{V}(\hat{d}) \le 2\max_{\mathbf{A}\in\mathcal{A}}\big\{\mathbb{E}[\delta^*(\mathbf{X}, \mathbf{A}) - \hat{\delta}(\mathbf{X}, \mathbf{A})]^{2}\big\}^{1/2}.
    \end{align}
    If Assumption \ref{margin_assumption} holds, the convergence rate is improved by
    \begin{align}
    \label{theorem1_bound2}
        \mathcal{V}(d^{*}) - \mathcal{V}(\hat{d}) \le C(\gamma)\max_{\mathbf{A}\in\mathcal{A}}\big\{\mathbb{E}[\delta^*(\mathbf{X}, \mathbf{A}) - \hat{\delta}(\mathbf{X}, \mathbf{A})]^{2}\big\}^{(1+\gamma)/(2+\gamma)},
    \end{align}
    where $C(\gamma)$ is a constant that depends on $C$ and $\gamma$.
\end{theorem}

Theorem \ref{theorem1} builds a connection between the value reduction and the estimation error of the treatment effects $\hat{\delta}(\cdot, \cdot)$, which shows that an accurate estimation of treatment effects would lead the estimated value function $\mathcal{V}(\hat{d})$ to approach the optimal value function $\mathcal{V}(d^{*})$. Based on Theorem \ref{theorem1}, we can further connect the value reduction bound to the excess risk of the estimator of the proposed model:

\begin{corollary} 
\label{corollary1}
  Suppose we define the expected risk of function $Q(\cdot, \cdot)$ as $L(Q) = \mathbb{E}[Y - Q(\mathbf{X}, \mathbf{A})]^2$. Then for any estimator of the function $Q(\cdot, \cdot)$, which is denoted by $\hat{Q}(\cdot, \cdot)$, we have the following value reduction bound:
    \begin{eqnarray}
        \mathcal{V}(d^{*}) - \mathcal{V}(\hat{d}) \le 2\big\{L(\hat{Q}) - L(Q^{*})\big\}^{1/2}. \notag
    \end{eqnarray}
    Further, if Assumption \ref{margin_assumption} holds, the above inequality can be tighter with $\gamma>0$:
    \begin{eqnarray}
        \mathcal{V}(d^{*}) - \mathcal{V}(\hat{d}) \le C(\gamma)\big\{L(\hat{Q}) - L(Q^{*})\big\}^{(1+\gamma)/(2+\gamma)}. \notag
    \end{eqnarray}
\end{corollary}

Next, we consider the value reduction bound under budget constraints. Since the multi-choice knapsack problem we formulated for budget-constrained ITR is NP-hard \citep{kellerer2004multidimensional}, we adopt a pseudo-polynomial dynamic programming algorithm \citep{dudzinski1987exact} to obtain an approximated solution. In the following, we analyze the theoretical property of the approximated value function that derived from dynamic programming Algorithm \ref{dp_for_constraints}. Specifically, we define the approximated value function as the sum of the treatment effects of the first $l$ subjects divided by the sample size $n$, which is $\hat{Z}_l(b)$ in Algorithm \ref{dp_for_constraints}. In addition, we denote the approximated value function as $Z_{l}^{*}(b)$ if the true treatment effects $\delta_{ij}^{*}$'s are plugged in. Then we have the following result indicating that the approximated value function converges if the estimation error of $\hat{\delta}(\cdot, \cdot)$ converges.

\begin{theorem}
\label{thm: value_reduction_constraint}
    For the approximated value function obtained from Algorithm \ref{dp_for_constraints}, for any $B > 0$, we have
    \begin{align}
        |Z^{*}(B) - \hat{Z}(B)| \le \frac{1}{n}\sum_{i=1}^n|\max_{\tilde{A}_{j}\in\mathcal{A}}\delta^*(\mathbf{x}_i, \tilde{A}_j) - \hat{\delta}(\mathbf{x}_i, \tilde{A}_j)|. \notag
    \end{align}
\end{theorem}

In other words, the approximated value function under budget constraints can converge if $\hat{\delta}(\cdot, \cdot)$ is a consistent estimator of treatment effects. Note that the proposed estimator is a doubly robust estimator in that either propensity score or treatment-free effects is correctly specified, our proposed estimator is a consistent estimator, which consequently leads the value function and approximated value function under budget constraints converge.

\subsection{Excess Risk Bound}

In this subsection, we provide a non-asymptotic value reduction bound for the proposed DEM and show the improved convergence rate under the DEM. In Corollary \ref{corollary1}, we have shown that the value reduction can be bounded by the excess risk between the true and estimated Q-functions. The excess risk serves as an intermediate tool to establish the non-asymptotic property of the proposed estimator which depends on the complexity of the function class. In the proposed method, we focus on the function class $\mathcal{Q} = \big\{Q: \mathcal{X}\times\mathcal{A}\rightarrow\mathbb{R} | Q(\mathbf{x}, \mathbf{a}) = m(\mathbf{x}) + \alpha(\mathbf{x})^T\beta(\mathbf{a})\big\}$, where $m(\cdot), \alpha(\cdot)$ and $\beta(\cdot)$ are defined in (\ref{treatment_free_function}), (\ref{cov_encoder}) and (\ref{trt_encoder}). We establish the following excess risk upper bound for the estimator in $\mathcal{Q}$: 

\begin{lemma}
\label{excess_risk_bound}
    For any distribution $(\mathbf{X}, \mathbf{A}, Y)$ with $\mathbb{E}[Y^2] \le c_1$, given a function $\hat{Q}$ from $\mathcal{Q}$, then with probability $1 - 2\epsilon$, 
    \begin{align}
    \label{lemma2_bound}
        L(\hat{Q}) - L(Q^{*}) \le 8C\mathcal{R}_{n}(\mathcal{Q}) + \sqrt{\frac{2c_1^2\log(1/\epsilon)}{n}},
    \end{align}
    where $C$ is the Lipschitz constant of $L(Q)$, and $\mathcal{R}_{n}(\mathcal{Q})$ is the Rademacher complexity of $\mathcal{Q}$.
\end{lemma}

Lemma \ref{excess_risk_bound} provides an upper bound of the excess risk in Corollary \ref{corollary1} using the Rademacher complexity of $\mathcal{Q}$. However, the Rademacher complexity of a general neural network is still an open problem in the literature and existing bounds are mainly established based on the different types of norm constraints of weight matrices \citep{bartlett2017spectrally,golowich2018size, neyshabur2017pac, neyshabur2015norm}. In this work, we focus on the following sub-class of $\mathcal{Q}$ with $L_2$ and spectral norm constraints:
\begin{eqnarray}
    \mathcal{Q}_{B_{m}, B_{\alpha}, B_{\beta}} = \big\{Q \in \mathcal{Q}: \lVert\mathbf{w}_{m}^{2}\rVert_2 \le B_{m}, \lVert\mathbf{W}_{m}^{1}\rVert_{2,\infty} \le B_{m}, \lVert\mathbf{T}_{l}\rVert_2 \le B_{\alpha}, \lVert\mathbf{U}_{l}\rVert_2 \le B_{\beta} \big\}, \notag
\end{eqnarray}

where $\lVert\cdot\rVert_2$ denotes the $L_2$-norm for vectors and the spectral norm for matrices. For any matrix $\mathbf{X} = (\mathbf{X}_1, ..., \mathbf{X}_p)$, and $\mathbf{X}_i$ is the $i$th column of matrix $\mathbf{X}$, we use $\lVert X\rVert_{2, \infty} = \max_{i}\lVert \mathbf{X}_i \rVert_2$ to denote the $L_{2,\infty}$ norm of $\mathbf{X}$. We then establish the upper bound of the Rademacher complexity of $\mathcal{Q}_{B_{m}, B_{\alpha}, B_{\beta}}$ as follows:

\begin{lemma}
\label{excess_risk_bound_2}
Suppose $\mathbb{E}[\lVert\mathbf{X}\rVert_2^2]\le c_2^2$. The Rademacher complexity of $\mathcal{Q}_{B_{m}, B_{\alpha}, B_{\beta}}$ is upper bounded by:
\begin{align}
    \label{lemma3_bound}
    \mathcal{R}_{n}(\mathcal{Q}_{B_{m}, B_{\alpha}, B_{\beta}}) &\le 2B_{m}^2c_2\sqrt{\frac{h}{n}} + B_{\alpha}^{L_{\alpha}}B_{\beta}^{L_{\beta}}c_2\sqrt{\frac{K}{n}}. 
\end{align}
\end{lemma}

Lemma \ref{excess_risk_bound_2} provides an upper bound of the Rademacher complexity of $\mathcal{F}_{B_{m}, B_{\alpha}, B_{\beta}}$ with the rate $O(\sqrt{\frac{1}{n}})$. The first term of (\ref{lemma3_bound}) is the upper bound for the function class of $m(\mathbf{x})$ in (\ref{treatment_free_function}), which depends on the width of hidden layers $h$. If $h$ is large, the function $m(\mathbf{x})$ is able to approximate a larger function space, but with a less tight upper bound on the generalization error. The second term of (\ref{lemma3_bound}) is associated with the functional class of the inner product of the double encoders with a convergence rate of $O(K^{1/2}n^{-1/2})$. The rate increases with the number of treatments $K$ rather than $|\mathcal{A}|$ due to the parameter-sharing feature of the interactive treatment encoder, and the linearly growing dimension of input of function $\beta(\cdot)$ in the proposed method. Specifically, the input of $\beta(\cdot)$ is the combination treatment $\mathbf{A}$ itself, and parameters in the treatment encoder are shared by all the combination treatments. Thus, the model complexity is proportional to $K$ and the product of the spectral norm of weight matrices. Based on Lemmas \ref{excess_risk_bound} and \ref{excess_risk_bound_2}, we derive the value reduction bound for the proposed method as follows:

\begin{theorem}
\label{theorem2}
For any distribution $(\mathbf{X}, \mathbf{A}, Y)$ with $\mathbb{E}[Y^2] \le c_1$, $\mathbb{E}[\lVert\mathbf{X}\rVert_2^2] \le c_2$. Consider the neural networks in the subspace $\mathcal{Q}_{B_{m}, B_{\alpha}, B_{\beta}}$, with probability at least $1-2\epsilon$, we have the following value reduction bound:
\begin{eqnarray}
\label{value_reduction_bound1}
    \mathcal{V}(d^{*}) - \mathcal{V}(\hat{d}) \le 2\bigg\{16CB_{m}^2c_2\sqrt{\frac{h}{n}} + 8CB_{\alpha}^{L_{\alpha}}B_{\beta}^{L_{\beta}}c_2\sqrt{\frac{K}{n}} + \sqrt{\frac{2c_1^2\log(1/\epsilon)}{n}}\bigg\}^{1/2}. \notag
\end{eqnarray}
If Assumption \ref{margin_assumption} holds, we have a tighter bound with a positive $\gamma$:
\begin{eqnarray}
\label{value_reduction_bound2}
    \mathcal{V}(d^{*}) - \mathcal{V}(\hat{d}) \le C(\gamma)\bigg\{16CB_{m}^2c_2\sqrt{\frac{h}{n}} + 8CB_{\alpha}^{L_{\alpha}}B_{\beta}^{L_{\beta}}c_2\sqrt{\frac{K}{n}} + \sqrt{\frac{2c_1^2\log(1/\epsilon)}{n}}\bigg\}^{(1+\gamma)/(2+\gamma)}. \notag
\end{eqnarray}
\end{theorem}

Theorem \ref{theorem2} establishes the value reduction bound in that the estimated decision rule can approach the optimal value function as the sample size increases. Compared with the existing value reduction bound for multi-arm treatments, the proposed method improves the convergence rate from $O(|\mathcal{A}|^{1/4})$ to $O((\log_{2}|\mathcal{A}|)^{1/4})$. Further, the order of the value reduction bound can approach nearly $n^{-1/2}$ as $\gamma$ goes to infinity, which is consistent with the convergence rates established in \citep{qian2011performance, qi2020multi}.

\section{Simulation Studies}
\label{sec: simulation}

In this section, we evaluate the performance of the proposed method in estimating the individualized treatment rule for combination treatments. Our numerical studies show that the proposed method achieves superior performance to competing methods in both unconstrained and budget-constrained scenarios.

\subsection{Unconstrained ITR simulation}
\label{sec: uncstr_simulation}

\begin{figure}
    \centering
    \begin{minipage}{0.45\textwidth}
    \includegraphics[width=1\textwidth]{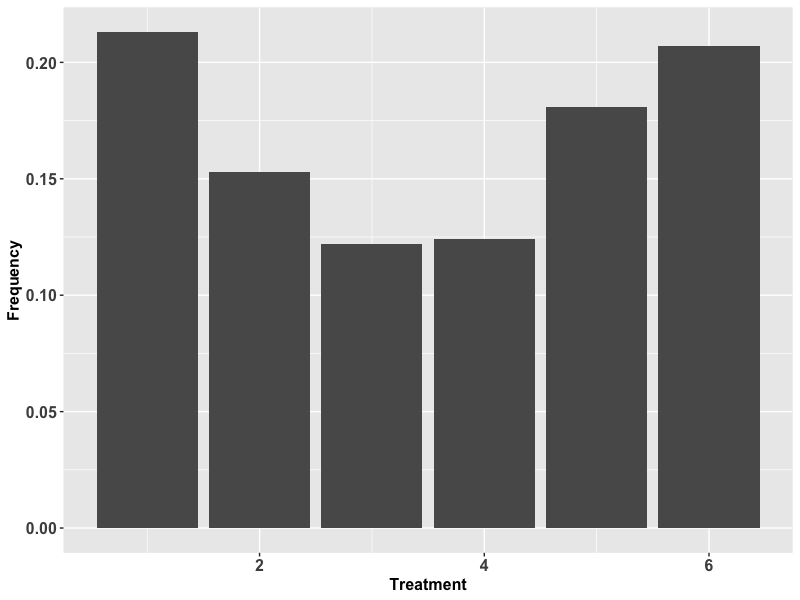}
    \end{minipage}
    \hfill
    \begin{minipage}{0.45\textwidth}
    \includegraphics[width=1\textwidth]{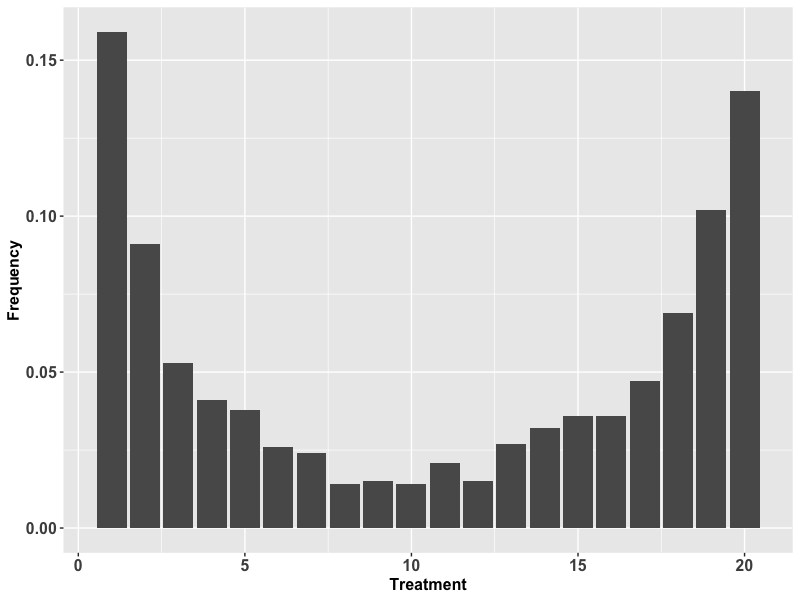}
    \end{minipage}
    \caption{Treatment assignment distribution in simulation settings. The left panel is for simulation settings 1 and 2, and the right panel is for simulation settings 3 and 4.}
    \label{fig: sim_asg}
\end{figure}

We first investigate the empirical performance of the proposed method without budget constraints. We assume the pre-treatment covariates $\mathbf{X} = (X_1, \ldots, X_{10}) \in \mathbb{R}^{10}$ are independently and uniformly sampled from $(-1, 1)$. Four simulation settings are designed to evaluate the performance under varying settings. In simulation settings 1 and 2, we consider combinations of 3 treatments, which induces 8 possible combinations, with 6 of them considered as our assigned treatments. Similarly, in simulation settings 3 and 4, we consider combinations of 5 treatments, and we assume that 20 of all combinations are assigned to subjects. The treatments are assigned either uniformly or following the propensity score model:
\begin{align}
\label{sim: propensity_score_model}
    \mathbb{P}(\tilde{A}_i|\mathbf{X}) = \frac{\exp\{0.2i * (\mathbf{X}^T\beta)\}}{\sum_{j}\exp\{0.2j * (\mathbf{X}^T\beta)\}},
\end{align}
and the marginal treatment assignment distribution is shown in Figure \ref{fig: sim_asg}.

\begin{table}
    \caption{\label{tab: sim_12_setting}Simulation settings 1 and 2: treatment effect and interaction effect functions specification. Column ``Treatment Effects'' specifies the treatment effect functions of individual treatments adopted in simulation settings 1 and 2. Column ``Interaction Effects'' specifies the interaction effects among individual treatments in setting 2.}
    \centering
    \begin{tabular}{p{20mm}p{60mm}p{60mm}}
    \hline
        Treatment $\mathbf{A}$ & Treatment Effects & Interaction Effects \\
    \hline
        $(0, 0, 0)$ & $0$ & -\\
        $(0, 0, 1)$ & $2X_{1} + \exp(X_{3} + X_{4})$ & - \\
        $(0, 1, 0)$ & $2X_{2}\log(X_{5}) + X_{7}$ & - \\
        $(0, 1, 1)$ & - & $\sin(5X_{1}^{2}) - 3(X_{2} - 0.5)^2$ \\
        $(1, 0, 0)$ & $\sin(X_{3}) + 2\log(X_{4}) + 2\log(X_{7})$ & -\\
        $(1, 1, 1)$ & - & $2\sin((X_{2} - X_{4})^2)$\\
    \hline
    \end{tabular}
\end{table}

\begin{table}
    \caption{\label{tab: sim_34_setting}Simulation settings 3 and 4: treatment effect and interaction effect functions specification. Column ``Treatment Effects'' specifies the treatment effect functions of individual treatments adopted in simulation settings 3 and 4. Column ``Interaction Effects'' specifies the interaction effects among individual treatments in setting 4.}
    \centering
    \begin{tabular}{p{20mm}p{60mm}p{60mm}}
    \hline
        Treatment $\mathbf{A}$ & Treatment Effects & Interaction Effects \\
    \hline
        $(0, 0, 0, 0, 0)$ & $0$ & -\\
        $(0, 0, 0, 0, 1)$ & $(X_{1} - 0.25) ^ 3$ & - \\
        $(0, 0, 0, 1, 0)$ & $2\log(X_{3}) + 4\log(X_{8})\cos(2\pi X_{10})$ & - \\
        $(0, 0, 1, 0, 0)$ & $X_{2}\sin(X_{4}) - 1$ &  \\
        $(0, 0, 1, 0, 1)$ & - & $\exp(2X_{2})$\\
        $(0, 1, 0, 0, 0)$ & $(X_{1} + X_{5} - X_{8} ^ 2)^3$ & -\\
        $(0, 1, 0, 0, 1)$ & - & $\exp(2X_{4} + X_{9})$\\
        $(0, 1, 0, 1, 1)$ & - & $-4\log(X_{6})$\\
        $(0, 1, 1, 0, 0)$ & - & $0$\\
        $(0, 1, 1, 1, 0)$ & - & $0$\\
        $(1, 0, 0, 0, 0)$ & $\exp(X_{2} - X_{5})$ & -\\
        $(1, 0, 0, 0, 1)$ & - & $0$ \\
        $(1, 0, 0, 1, 0)$ & - & $0$ \\
        $(1, 0, 1, 0, 0)$ & - & $0$ \\
        $(1, 0, 1, 0, 1)$ & - & $-3/2\cos(2\pi X_{1} + X_{8}^2)$\\
        $(1, 1, 0, 0, 0)$ & - & $0$\\
        $(1, 1, 0, 0, 1)$ & - & $-4\log(X_{6})$\\
        $(1, 1, 0, 1, 1)$ & - & $X_{6}^2 + 1/2\sin(2\pi/X_{7})$\\
        $(1, 1, 1, 0, 0)$ & - & $0$\\
        $(1, 1, 1, 1, 0)$ & - & $0$\\
    \hline
    \end{tabular}
\end{table}

In simulation setting 1, we assume that the treatment effects of the combination treatment are additive from individual treatment, and we specify the individual treatment effect functions in the column ``Treatment Effects'' provided in Table \ref{tab: sim_12_setting}. Based on simulation setting 1, we consider some interaction effects among treatments in simulation setting 2, which are specified in the column ``Interaction Effects'' in Table \ref{tab: sim_12_setting}. Therefore, the treatment effects of the combination treatments are the summation of individual treatment effects and interaction effects. Similarly, Table \ref{tab: sim_34_setting} specifies the treatment effects and interaction effects for simulation settings 3 and 4 in the same manner. In particular, the treatment effects of the combination treatments are additive from individual treatment effects in the simulation setting 3, while interaction effects are added in the simulation setting 4. In summary, we evaluate the empirical performance of the proposed method and competing methods under the additive treatment effects scenarios in simulation settings 1 and 3, and under the interactive treatment effects scenarios in simulation settings 2 and 4.

For each simulation setting, the sample sizes for the training data vary from 500, 1000 to 2000, and each setting is repeated 200 times. Then we compare the proposed method with the following methods: the $L_1$-penalized least square ($L_1$-pls, \citealt{qian2011performance}), the outcome weighted learning with multinomial deviance (OWL-MD, \citealt{huang2019multicategory}), the multicategory outcome weighted learning with linear decisions(MOWL-linear \citealt{zhang2014multicategory}), the outcome weighted learning with deep learning (OWL-DL, \citealt{liang2018estimating}), the treatment-agnostic representation network (TARNet) \citep{shalit2017estimating}. The empirical evaluation of the value function and accuracy are reported in Tables \ref{tab: sim1_value} and \ref{tab: sim1_accuracy}, where the empirical value function \citep{qian2011performance} is calculated via
\begin{eqnarray}
    \hat{\mathcal{V}}(d) = \frac{\mathbb{E}_{n}[YI\{d(\mathbf{X}) = \mathbf{A}\}]}{\mathbb{E}_{n}[I\{d(\mathbf{X}) = \mathbf{A}\}]}, \notag
\end{eqnarray}
where $\mathbb{E}_{n}$ denotes the empirical average. 

Since simulation settings 1 and 3 do not include interaction effects among different treatments, all competing methods except for the OWL-DL \citep{liang2018estimating} are over-parameterized, while the proposed method can be adaptive to the additive setting with a large $\lambda_{i}$. Therefore, the proposed method and the OWL-DL outperform other competing methods in both settings. In contrast, complex interaction effects are considered in simulation settings 2 and 4, and the performance of OWL-DL is inferior since a consistent estimation is not guaranteed for OWL-DL if there are interaction effects. Although other competing methods are saturated in incorporating interaction effects, their estimation efficiencies are still undermined since the decision functions in these methods are all treatment-specific, while the proposed method possesses the unique parameter-sharing feature for different treatments. Therefore, the advantage of our method is more significant for small sample sizes or large $K$ scenarios. Specifically, the proposed method improves the accuracy by 10.9\% to 17.8\% in simulation setting 4 when the sample size is 500. In addition, we also compare the empirical performance of the double encoder model with different choices of covariates and treatment encoders, and the detailed simulation results are presented in the supplementary materials.

\begin{table}
\caption{\label{tab: sim1_value}Unconstrained simulation study: Comparisons of value functions for the proposed method and existing methods including the $L_1$-penalized least square ($L_1$-pls, \citealt{qian2011performance}), the outcome weighted learning with multinomial deviance (OWL-MD, \citealt{huang2019multicategory}), the multicategory outcome weighted learning with linear decisions (MOWL-linear, \citealt{zhang2020multicategory}), the outcome weighted learning with deep learning (OWL-DL, \citealt{liang2018estimating}) and the treatment-agnostic representation network (TARNet, \citealt{shalit2017estimating}). Two treatment assignment schemes are presented: all treatments are uniformly assigned to subjects (Uniform), and treatments are assigned based on the propensity score model (\ref{sim: propensity_score_model}, PS-based).}
\centering
\scriptsize
\begin{tabular}{p{15mm}p{12mm}p{10mm}p{16mm}p{16mm}p{16mm}p{16mm}p{16mm}p{16mm}p{16mm}}
\hline
\multicolumn{8}{c}{Value} \\
\hline
Treatment Assignment & Setting & Sample Size & \textbf{Proposed} & $L_1$-pls  & OWL-MD  & MOWL-linear  & OWL-DL & TARNet\\
\hline
\multirow{12}{0pt}{Uniform}&\multirow{3}{0pt}{1} & 500 & \textbf{5.477(0.218)} & 3.643(0.158) & 5.338(0.259) & 5.206(0.185) & 5.408(0.265) & 5.320(0.278)\\
& & 1000 & \textbf{5.622(0.206)} & 3.800(0.154) & 5.510(0.251) & 5.292(0.182) & 5.512(0.259) & 5.398(0.273)\\
& & 2000 & \textbf{5.658(0.208)} & 3.989(0.149) & 5.577(0.246) & 5.384(0.179) & 5.595(0.251) & 5.411(0.272)\\ \cline{3-9}
&\multirow{3}{0pt}{2} & 500 & \textbf{5.268(0.325)} & 3.870(0.291) & 5.132(0.306) & 5.078(0.291) & 5.015(0.400) & 5.008(0.405)\\
& & 1000 & \textbf{5.418(0.312)} & 4.028(0.285) & 5.302(0.299) & 5.168(0.279) & 5.105(0.398) & 5.024(0.402)\\
& & 2000 & \textbf{5.498(0.311)} & 4.191(0.282) & 5.344(0.289) & 5.211(0.272) & 5.215(0.382) & 5.118(0.386)\\ \cline{3-9}
&\multirow{3}{0pt}{3} & 500 & \textbf{5.600(0.288)} & 3.562(0.235) & 4.479(0.285) & 5.216(0.240) & 5.332(0.312) & 5.354(0.305)\\
& & 1000 & \textbf{5.667(0.268)} & 3.702(0.232) & 4.987(0.279) & 5.283(0.241) & 5.403(0.310) & 5.366(0.300)\\
& & 2000 & \textbf{5.719(0.262)} & 3.855(0.230) & 5.274(0.265) & 5.459(0.232) & 5.598(0.299) & 5.423(0.289)\\ \cline{3-9}
&\multirow{3}{0pt}{4} & 500 & \textbf{6.117(0.328)} & 4.490(0.264) & 5.900(0.278) & 5.948(0.277) & 5.995(0.335) & 5.895(0.328)\\
& & 1000 & \textbf{6.374(0.319)} & 4.850(0.262) & 6.200(0.272) & 6.120(0.269) & 6.012(0.321) & 5.998(0.315)\\
& & 2000 & \textbf{6.732(0.310)} & 5.252(0.254) & 6.506(0.259) & 6.494(0.262) & 6.254(0.311) & 6.057(0.310)\\ \cline{3-9}
\hline
\multirow{12}{0pt}{PS-based}&\multirow{3}{0pt}{1} & 500 & \textbf{5.415(0.238)} & 4.048(0.198) & 5.061(0.215) & 4.897(0.233) & 5.218(0.273) & 5.013(0.305)\\
& & 1000 & \textbf{5.589(0.223)} & 4.087(0.198) & 5.099(0.213) & 4.959(0.231) & 5.223(0.272) & 5.018(0.298)\\
& & 2000 & \textbf{5.662(0.219)} & 4.224(0.183) & 5.178(0.201) & 4.898(0.230) & 5.238(0.279) & 5.017(0.299)\\ \cline{3-9}
&\multirow{3}{0pt}{2} & 500 & \textbf{5.005(0.324)} & 3.980(0.236) & 4.815(0.254) & 4.629(0.279) & 4.635(0.336) & 4.886(0.352)\\
& & 1000 & \textbf{5.622(0.322)} & 4.042(0.235) & 4.906(0.249) & 4.700(0.276) & 4.913(0.334) & 5.021(0.341)\\
& & 2000 & \textbf{5.658(0.320)} & 4.104(0.229) & 5.005(0.245) & 4.672(0.274) & 4.998(0.326) & 5.054(0.331)\\ \cline{3-9}
&\multirow{3}{0pt}{3} & 500 & \textbf{5.665(0.330)} & 3.384(0.258) & 3.540(0.269) & 5.401(0.302) & 5.505(0.352) & 5.476(0.338)\\
& & 1000 & \textbf{5.792(0.321)} & 4.560(0.248) & 5.009(0.268) & 5.519(0.303) & 5.784(0.348) & 5.676(0.308)\\
& & 2000 & \textbf{5.796(0.318)} & 5.273(0.246) & 5.307(0.259) & 5.582(0.300) & 5.788(0.338) & 5.774(0.299)\\ \cline{3-9}
&\multirow{3}{0pt}{4} & 500 & \textbf{5.630(0.356)} & 4.462(0.285) & 3.816(0.305) & 5.090(0.321) & 5.028(0.405) & 5.108(0.387) \\
& & 1000 & \textbf{6.001(0.348)} & 5.432(0.355) & 5.822(0.302) & 5.134(0.319) & 5.384(0.400) & 5.338(0.379)\\
& & 2000 & \textbf{6.289(0.345)} & 5.808(0.344) & 6.141(0.299) & 5.294(0.318) & 5.684(0.389) & 5.589(0.378)\\
\hline
\end{tabular}
\end{table}

\begin{table}
\caption{\label{tab: sim1_accuracy}Unconstrained simulation study: Comparisons of accuracies for the proposed method and existing methods including the $L_1$-penalized least square ($L_1$-pls, \citealt{qian2011performance}), the outcome weighted learning with multinomial deviance (OWL-MD, \citealt{huang2019multicategory}), the multicategory outcome weighted learning with linear decisions (MOWL-linear, \citealt{zhang2020multicategory}), the outcome weighted learning with deep learning (OWL-DL, \citealt{liang2018estimating}) and the treatment-agnostic representation network (TARNet, \citealt{shalit2017estimating}). Two treatment assignment schemes are presented: all treatments are uniformly assigned to subjects (Uniform), and treatments are assigned based on the propensity score model (\ref{sim: propensity_score_model}, PS-based).}
\centering
\scriptsize
\begin{tabular}{p{15mm}p{12mm}p{10mm}p{16mm}p{16mm}p{16mm}p{16mm}p{16mm}p{16mm}p{16mm}}
\hline
\multicolumn{8}{c}{Accuracy} \\
\hline
Treatment Assignment & Setting & Sample Size & \textbf{Proposed} & $L_1$-pls  & OWL-MD  & MOWL-linear  & OWL-DL & TARNet\\
\hline
\multirow{12}{0pt}{Uniform}&\multirow{3}{0pt}{1} & 500 & \textbf{0.622(0.052)} & 0.338(0.039) & 0.572(0.040) & 0.491(0.041) & 0.598(0.046) & 0.553(0.050)\\
& & 1000 & \textbf{0.707(0.050)} & 0.358(0.038) & 0.642(0.040) & 0.507(0.040) & 0.682(0.043) & 0.640(0.050) \\
& & 2000 & \textbf{0.738(0.050)} & 0.382(0.038) & 0.694(0.039) & 0.552(0.042) & 0.710(0.042) & 0.686(0.048) \\ \cline{3-9}
&\multirow{3}{0pt}{2} & 500 & \textbf{0.544(0.058)} &0.350(0.037) & 0.515(0.041) & 0.474(0.035) & 0.488(0.045) & 0.432(0.052)\\
& & 1000 & \textbf{0.610(0.054)} & 0.367(0.034) & 0.573(0.038) & 0.505(0.033) & 0.532(0.040) & 0.462(0.051)\\
& & 2000 & \textbf{0.630(0.051)} & 0.389(0.033) & 0.605(0.036) & 0.516(0.034) & 0.568(0.041) & 0.506(0.049)\\ \cline{3-9}
&\multirow{3}{0pt}{3} & 500 & \textbf{0.420(0.041)} & 0.102(0.025) & 0.148(0.031) & 0.251(0.034) & 0.271(0.047) & 0.205(0.057)\\
& & 1000 & \textbf{0.445(0.039)} & 0.099(0.024) & 0.187(0.027) & 0.285(0.033) & 0.300(0.046) & 0.268(0.053)\\
& & 2000 & \textbf{0.464(0.039)} & 0.116(0.021) & 0.254(0.027) & 0.331(0.031) & 0.353(0.048) & 0.311(0.045)\\ \cline{3-9}
&\multirow{3}{0pt}{4} & 500 & \textbf{0.324(0.045)} & 0.146(0.031) & 0.205(0.032) & 0.215(0.032) & 0.154(0.047) & 0.162(0.041)\\
& & 1000 & \textbf{0.335(0.044)} & 0.191(0.031) & 0.279(0.030) & 0.229(0.032) & 0.189(0.045) & 0.193(0.043)\\
& & 2000 & \textbf{0.372(0.041)} & 0.222(0.029) & 0.323(0.029) & 0.241(0.029) & 0.228(0.044) & 0.225(0.042)\\ \cline{3-9}
\hline
\multirow{12}{0pt}{PS-based}&\multirow{3}{0pt}{1} & 500 & \textbf{0.571(0.048)} & 0.317(0.031) & 0.477(0.035) & 0.430(0.037) & 0.498(0.050) & 0.432(0.053)\\
& & 1000 & \textbf{0.648(0.044)} & 0.327(0.027) & 0.502(0.033) & 0.451(0.032) & 0.525(0.047) & 0.462(0.051)\\
& & 2000 & \textbf{0.699(0.044)} & 0.332(0.028) & 0.522(0.031) & 0.448(0.031) & 0.552(0.048) & 0.481(0.050)\\ \cline{3-9}
&\multirow{3}{0pt}{2} & 500 & \textbf{0.492(0.052)} & 0.284(0.037) & 0.420(0.037) & 0.397(0.036) & 0.370(0.047) & 0.375(0.048)\\
& & 1000 & \textbf{0.566(0.051)} & 0.290(0.037) & 0.443(0.034) & 0.414(0.037) & 0.397(0.045) & 0.388(0.047)\\
& & 2000 & \textbf{0.618(0.048)} & 0.291(0.037) & 0.464(0.031) & 0.406(0.032) & 0.411(0.041) & 0.425(0.049)\\ \cline{3-9}
&\multirow{3}{0pt}{3} & 500 & \textbf{0.378(0.041)} & 0.053(0.030) & 0.071(0.033) & 0.223(0.036) & 0.300(0.051) & 0.248(0.050)\\
& & 1000 & \textbf{0.439(0.041)} & 0.091(0.031) & 0.181(0.032) & 0.279(0.034) & 0.376(0.041) & 0.344(0.048)\\
& & 2000 & \textbf{0.444(0.040)} & 0.137(0.029) & 0.242(0.030) & 0.327(0.031) & 0.416(0.038) & 0.378(0.047)\\ \cline{3-9}
&\multirow{3}{0pt}{4} & 500 & \textbf{0.223(0.056)} & 0.101(0.038) & 0.083(0.039) & 0.090(0.043) & 0.102(0.052) & 0.084(0.061)\\
& & 1000 & \textbf{0.267(0.053)} & 0.136(0.036) & 0.195(0.041) & 0.089(0.041) & 0.121(0.051) & 0.098(0.056)\\
& & 2000 & \textbf{0.279(0.052)} & 0.205(0.037) & 0.245(0.041) & 0.101(0.039) & 0.168(0.048) & 0.127(0.055)\\
\hline
\end{tabular}
\end{table}

\newpage
\subsection{Budget-constrained ITR simulation}
\label{sec: cstr_simulation}

\begin{table}
\caption{\label{tab: sim2_value}Constrained simulation study: Comparisons of value functions for the proposed method and existing methods including the $L_1$-penalized least square ($L_1$-pls, \citealt{qian2011performance}), the outcome weighted learning with multinomial deviance (OWL-MD, \citealt{huang2019multicategory}), the multicategory outcome weighted learning with linear decisions (MOWL-linear, \citealt{zhang2020multicategory}), and the treatment-agnostic representation network (TARNet, \citealt{shalit2017estimating}). All treatments are uniformly assigned to subjects.}
\centering
\scriptsize
\begin{tabular}{p{12mm}p{12mm}p{16mm}p{16mm}p{16mm}p{16mm}p{16mm}}
\hline
\multicolumn{7}{c}{Value} \\
\hline
Sample Size & Constraints & \textbf{Proposed} & $L_1$-pls  & OWL-MD  & MOWL-linear  & TARNet\\
\hline
\multirow{4}{0pt}{500} & 20\% & \textbf{5.237(0.322)} & 3.445(0.268) & 4.918(0.283) & 4.909(0.279) & 4.874(0.335)\\
& 50\% & \textbf{5.527(0.320)} & 3.756(0.277) & 5.218(0.271) & 5.215(0.273) & 4.998(0.335)\\
& 80\% & \textbf{5.832(0.319)} & 4.108(0.267) & 5.510(0.280) & 5.499(0.276) & 5.425(0.329)\\
& 100\% & \textbf{6.117(0.328)} & 4.490(0.264) & 5.900(0.278) & 5.948(0.277) & 5.895(0.328)\\ 
\hline
\multirow{4}{0pt}{1000} & 20\% & \textbf{5.498(0.321)} & 3.685(0.261) & 5.175(0.270) & 5.170(0.271) & 5.047(0.320)\\
& 50\% & \textbf{5.841(0.315)} & 4.014(0.265) & 5.487(0.269) & 5.318(0.275) & 5.274(0.318)\\
& 80\% & \textbf{6.102(0.320)} & 4.417(0.261) & 5.612(0.273) & 5.598(0.267) & 5.418(0.315)\\
& 100\% & \textbf{6.374(0.319)} & 4.850(0.262) & 6.200(0.272) & 6.120(0.269) & 5.998(0.315)\\ 
\hline
\multirow{4}{0pt}{2000} & 20\% & \textbf{5.789(0.309)} & 4.108(0.249) & 5.356(0.209) & 5.317(0.266) &  5.015(0.308)\\
& 50\% & \textbf{6.015(0.315)} & 4.437(0.251) & 5.897(0.207) & 5.778(0.261) & 5.298(0.298)\\
& 80\% & \textbf{6.324(0.311)} & 4.847(0.255) & 6.215(0.208) & 6.117(0.264) & 5.598(0.315)\\
& 100\% & \textbf{6.732(0.310)} & 5.252(0.254) & 6.506(0.201) & 6.494(0.262) & 6.057(0.310)\\
\hline
\end{tabular}
\end{table}

In this subsection, we investigate the budget-constrained setting with the same data generation mechanism as in simulation setting 4 in Section \ref{sec: uncstr_simulation}. For a fair comparison, we compare the proposed method with competing methods which provide a score to measure the utility or effect of each treatment. We then apply the proposed MCKP method to all of these methods because the proposed framework (\ref{empirical_constrained_problem}) does not require specification for the treatment effects.

For the budget constraint, we let the second treatment be the most critical, or urgently needed, by the population. Thus, we constrain the amount of the second treatment so that only partial patients can be treated by the second treatment. The quantiles of the constrained populations are $20\%, 50\%, 80\%$, and $100\%$, where the constraints in the last case is trivial constraints as it is equivalent to an unconstrained setting. 

The simulation results are provided in Tables \ref{tab: sim2_value}, which clearly indicate that the proposed method outperforms other methods in the constrained cases. Moreover, the proposed method achieves smaller reductions of value functions when budget constraints are imposed. Specifically, the value functions of the competing methods are reduced by about 0.9 when the budget is decreased from 100$\%$ to 20$\%$. More precisely, when the sample size is 2000, compared with the best performances of competing methods, the proposed method improves the value function by $8.08\%$ when the budget is $20\%$, and it achieves $3.47\%$ improvement in value function when the budget is $100\%$. The significant improvement of the value function under limited budget scenarios shows that the proposed method provides a more accurate estimation of treatment effects, and thus leads to better individualized treatment rule estimation under restrictive constraints.

\section{Application to patient-derived xenograft study}
\label{sec: pdx}

\begin{figure}[t]
    \centering
    \includegraphics[width=1.0\textwidth]{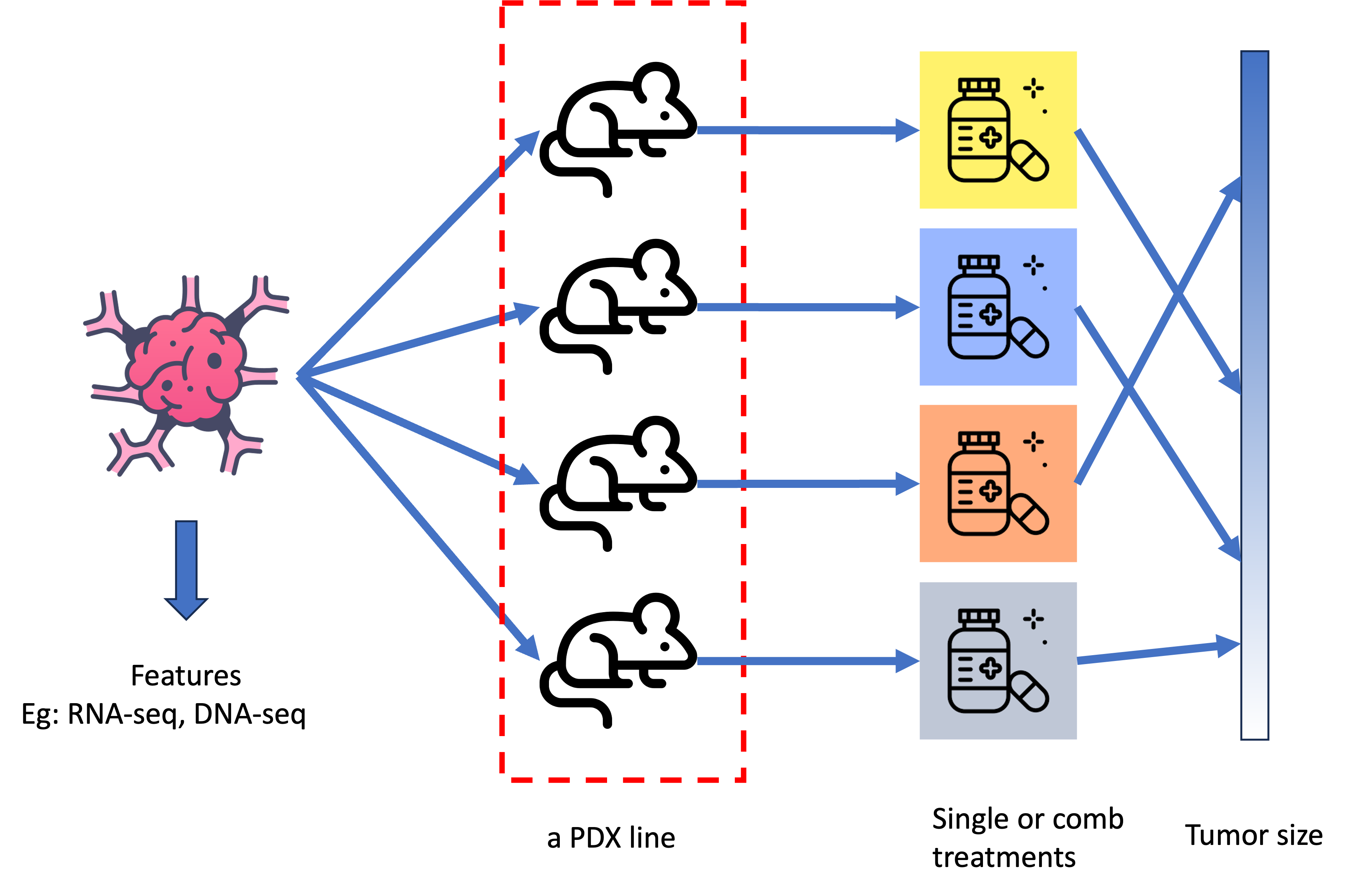}
    \caption{Illustration of the PDX data collection. Tumor samples from a patient are implanted into multiple mice, where a PDX line is formed by these mice. Different treatments can be applied simultaneously. Tumor size, which is the primary interest of the outcome, is measured for each mouse. RNA and DNA sequencing and other features are collected as pre-treatment covariates.}
    \label{fig: pdx_study}
\end{figure}

In this section, we apply our method to patient-derived xenograft (PDX) data to inform optimal personalized treatment for cancer patients. Due to ethical issues and other limitations of randomized clinical trials in cancer studies, recent works have utilized patient-derived xenografts (PDXs) to perform large-scale screening in mice to evaluate cancer therapies \citep{gao2015high}. Specifically, samples of primary solid tumors are collected from patients through surgery or biopsy \citep{hidalgo2014patient}, and each tumor is implanted into multiple mice to create a PDX line, where multiple treatments can be applied simultaneously. Meanwhile, high throughput genomic assays such as RNA and DNA sequencing of solid tumors are collected as the pre-treatment covariates. Therefore, the mice within one PDX line share the same pre-treatment covariates. The primary interest of the outcome in these studies is the tumor size or growth rate. An illustration of the PDX data collection procedure is provided in Figure \ref{fig: pdx_study}.

The original data is from the Novartis PDX study \citep{gao2015high}, and we follow the screening and pre-processing steps in \citep{rashid2020high} to obtain the genomic data and tumor measurements. In particular, our focus is colorectal cancer (CRC), where 37 PDX lines are investigated and the 13 treatments listed in Table \ref{tab: pdx_trt} are administered simultaneously to all 37 PDX lines. These lead to 481 observations. Compared with other randomized clinical trials or observational studies, the outcomes of each subject (PDX line) in PDX data given all treatments are fully observed. Therefore, the positivity assumption is intrinsically satisfied. In addition, the pre-treatment covariates are 94 genetic features associated with 50 genes selected by unsupervised and supervised screening by \citep{rashid2020high} and these pre-treatment covariates are inherently balanced since all treatment groups include exactly the same PDX lines. Furthermore, the outcome of our interest is measured by the scaled maximum observed tumor size shrunken from the baseline size, where a larger value is more desirable.

For the budget-constrained setting, we impose the costs of treatments as follows: \$79 for BKM120, \$100 for LJC049, \$66 for BYL719, \$240 for cetuximab, \$124 for encorafenib, \$500 for LJM716 and \$79 for binimetinib, where the prices for LJC049 and LJM716 are hypothetical, while other prices are based on \url{https://www.goodrx.com} for unit dosage. We consider the hypothetical budgets for these 37 PDX lines as \$21,000, \$15,000, \$10,000, and \$5,000, where \$21,000 is equivalent to the unconstrained scenario because it is sufficient to cover the most expensive combination treatment for all PDX lines. 

To implement our proposed method and other competing methods, we randomly split the dataset into training (25 PDX lines), validation (6 PDX lines), and testing (6 PDX lines) sets. All methods are trained on the training set, while hyper-parameters are tuned based on the validation set. The value function of the unconstrained scenario is calculated on testing set via $\hat{\mathcal{V}}(d) = \frac{\mathbb{E}_{n}[YI\{d(\mathbf{X}) = \mathbf{A}\}]}{\mathbb{E}_{n}[I\{d(\mathbf{X}) = \mathbf{A}\}]}$. For the budget-constrained scenarios, we apply the selected model to all PDX lines to obtain the estimation of treatment effects, and then apply the MCKP algorithm to all PDX lines and calculate the value function based on 37 PDX lines. Finally, we repeat the above random splitting 100 times to validate the results for comparison.

\begin{table}
    \caption{\label{tab: pdx_trt}Table of single or combination treatments considered to treat colorectal cancer for PDX lines.}
    \centering
    \begin{tabular}{p{10mm}p{15mm}p{15mm}p{15mm}p{15mm}p{15mm}p{15mm}p{15mm}}
    \hline
    Type & BKM120 & LJC049 & BYL719 & cetuximab & encorafenib & LJM716 & binimetinib \\
    \hline
    Single & 1 & 0 & 0 & 0 & 0 & 0 & 0\\ 
    Single & 0 & 1 & 0 & 0 & 0 & 0 & 0\\
    Single & 0 & 0 & 1 & 0 & 0 & 0 & 0\\
    Single & 0 & 0 & 0 & 1 & 0 & 0 & 0\\
    Single & 0 & 0 & 0 & 0 & 1 & 0 & 0\\
    Single & 0 & 0 & 0 & 0 & 0 & 0 & 1\\
    Comb & 1 & 1 & 0 & 0 & 0 & 0 & 0\\
    Comb & 0 & 0 & 1 & 1 & 0 & 0 & 0\\
    Comb & 0 & 0 & 1 & 1 & 1 & 0 & 0\\
    Comb & 0 & 0 & 1 & 0 & 1 & 0 & 0\\
    Comb & 0 & 0 & 1 & 0 & 0 & 1 & 0\\
    Comb & 0 & 0 & 1 & 0 & 0 & 0 & 1\\
    Comb & 0 & 0 & 0 & 1 & 1 & 0 & 0\\
    \hline
    \end{tabular}
\end{table}

\begin{table}
\caption{\label{tab: real_data}Mean and standard errors of value function under different budget constraints. The budget \$21,000 is equivalent to the non-constrained scenario.}
\centering
{\scriptsize
    \begin{tabular}{p{15mm}p{20mm}p{20mm}p{20mm}p{20mm}p{20mm}p{20mm}p{20mm}}
    \hline
    Budgets & Proposed method & $L_1$-PLS & MOWL-linear & MOWL-kernel & OWL-DL & TARNet \\
    \hline
    \$ 21,000 & \textbf{0.142(0.199)} & -0.157(0.227) & 0.054(0.205) & -0.313(0.251) & -0.298(0.268) & -0.254(0.209) \\
    \hline
    \$ 15,000 & \textbf{0.103(0.196)} & -0.188(0.231) & -0.007(0.207) & -0.312(0.255) & -0.318(0.264) & -0.283(0.210) \\
    \$ 10,000 & \textbf{0.067(0.200)} & -0.209(0.222) & -0.077(0.198) & -0.330(0.250) & -0.320(0.267) & -0.299(0.208) \\
    \$ 5,000 & \textbf{0.050(0.202)} & -0.282(0.216) & -0.103(0.199) & -0.367(0.249) & -0.367(0.259) & -0.348(0.207) \\
    \hline
    \end{tabular}
}
\end{table}

In the following, we report the means and standard deviations of the value functions under different budget constraints in Table \ref{tab: real_data}. For the unconstrained scenario ($\$21,000$ budget), the proposed method achieves great improvement in value maximization. As a reference, the value function under one-size-fits-all rules and the optimal treatment are shown in Table \ref{tab: real_data_value_universal}, which shows that our proposed method improves the value function gap between optimal treatment assignment and one-size-fits-all rules by 23.0\% to 81.1\%. In comparison with the competing methods which also estimate the individualized treatment rule, our proposed method improves the value function gap between the optimal treatment assignment and competing ITR methods by 40.4\% to 65.8\%. 

For the budget-constrained scenarios, the proposed method achieves dominant advantages over the competing methods in that the value function under the most restrictive budget constraints still achieves comparable results with other competing methods in the unconstrained scenario. Compared with one-size-fits-all rules, the proposed method achieves the best value function with about \$5,000 budget compared to the best one-size-fits-all rule, which is the combination of BYL719 and binimetinib with a \$5,365 budget. In summary, the proposed method is more capable of effectively controlling the tumor size than any other competing methods and one-size-fits-all rules. Our approach could have great potential for improving therapy quality for colorectal cancer patients.

\begin{table}
    \caption{\label{tab: real_data_value_universal}Value function under one-size-fits-all rules and optimal treatment assignment}
    \centering
    \begin{tabular}{p{60mm}p{10mm}p{10mm}|p{20mm}p{10mm}p{10mm}}
    \hline
    Treatment & Value & Budget & Treatment & Value & Budget\\
    \hline
    BKM120 + LJC049 & -0.684 & \$6,623 & BKM120 & -0.350 & \$2,923 \\
    BYL719 + cetuximab & -0.232 & \$11,322 & LJC049 & -1.168 & \$3,700\\
    BYL719 + cetuximab + encorafenib & -0.388 & \$15,910 & BYL719 & -0.371 & \$2,442 \\
    BYL719 + encorafenib & -0.562 & \$7,030 & cetuximab & -0.633 & \$8,880 \\
    BYL719 + LJM716& -0.103 & \$20,942 & encorafenib & -1.083 & \$4,588 \\
    \textbf{BYL719 + binimetinib} & \textbf{0.047} & \textbf{\$5,365} & binimetinib & -0.425 & \$2,923\\
    cetuximab + encorafenib & -0.749 & \$13,468 & \textit{Optimal} & 0.447 & \$8,171 \\
    \hline
    \end{tabular}
\end{table}

\section{Discussion}
In this paper, we broaden the scope of estimating the individualized treatment rule (ITR) from binary and multi-arm treatments to combination treatments, where treatments within each combination can interact with each other. We propose the Double Encoder Model (DEM) as a nonparametric approach to accommodate intricate treatment effects of combination treatments. Specifically, our method overcomes the curse of dimensionality issue via adopting neural network treatment encoders. The parameter-sharing feature of the neural network treatment encoder enhances the estimation efficiency such that the proposed method is able to outperform other parametric approaches given a small sample size. In addition, we also adapt the estimated ITR to budget-constrained scenarios. This adapation is achieved through the multi-choice knapsack framework, which strengthens our proposed method in situations with limited resources. Theoretically, we offer a value reduction bound with and without budget constraints and an improved convergence rate concerning the number of treatments under the DEM.

Several potential research directions could be worth exploring further. First of all, the proposed method employs the propensity score model to achieve the double robustness property. However, the inverse probability weighting method could be weakened in observational studies considering the combination treatments, due to the potential violation of positivity assumptions. This phenomenon is also observed in the binary treatment scenario with high-dimensional covariates \citep{d2021overlap}. There are  existing works to overcome this limitation in the binary and multi-arm treatment setting, which utilizes overlap weights\citep{li2019propensity, lifan2019addressing} to substitute the propensity score. However, this strategy cannot solve the same issue in combination treatment problems. Therefore, exploring alternative approaches for combination treatment problems will be a worthwhile direction.

Second, compared with the binary treatments, combination treatments enable us to optimize multiple outcomes of interest simultaneously. The major challenge of multiple outcomes is that each combination treatment may only favor a few outcomes, and therefore an optimal ITR is expected to achieve a trade-off among multiple outcomes. Some recent works have studied trade-offs between the outcome of interest and risk factors \citep{wang2018learning, huang2020estimating}. However, trade-offs among multiple outcomes could be more challenging.

Furthermore, interpretability is another desirable property of the individualized treatment rule, especially in medical settings. The proposed method incorporates neural networks to enjoy benefits in estimation and theoretical properties, while interpretation is not obvious. Some existing works \citep{laber2015tree, zhang2015using} propose tree-type methods for better interpretability under binary or multi-arm settings, but may be not applicable for combination treatments. In the literature of explainable machine learning, there are available post-hoc and model-agnostic approaches \citep{lundberg2017unified, shrikumar2017learning} which can be learned from. However, more sophisticated adaptation might be needed for the combination treatment problem.

\bibliographystyle{rss}
\bibliography{reference}

\end{document}